\documentclass[times,twocolumn,final]{elsarticle}

\usepackage{medima}
\usepackage{framed,multirow}

\usepackage{amssymb}
\usepackage{latexsym}

\usepackage{lineno}
\usepackage{subfig}
\usepackage{float}
\usepackage{amsmath}
\usepackage{graphicx}
\usepackage{multirow}
\usepackage{amsmath}
\usepackage{booktabs}
\usepackage{makecell}
\setcellgapes{2pt}
\usepackage{setspace}
\usepackage{adjustbox}
\usepackage{csquotes}
\usepackage[normalem]{ulem}
\usepackage{textcomp}
\usepackage{xcolor}
\usepackage{threeparttable}
\usepackage{amsmath,amssymb,amsfonts}
\usepackage{graphicx}
\usepackage{textcomp}
\usepackage{amssymb}
\usepackage{graphicx}
\usepackage{amsmath}
\usepackage{commath}
\usepackage{mathtools}
\usepackage{comment}
\usepackage{picins}
\usepackage[linesnumbered,ruled,lined,boxed]{algorithm2e}
\newcommand{\lyxmathsym}[1]{\ifmmode\begingroup\def\b@ld{bold}
	\text{\ifx\math@version\b@ld\bfseries\fi#1}\endgroup\else#1\fi}

\usepackage{amstext}

\usepackage{array}

\usepackage{amssymb}
\floatstyle{ruled}

\usepackage{graphicx}
\usepackage{amsmath}
\usepackage{commath}
\usepackage{mathtools}

\usepackage{url}
\usepackage{xcolor}

\usepackage{hyperref}

\definecolor{newcolor}{rgb}{.8,.349,.1}

\journal{Medical Image Analysis}

\begin{document}

\verso{Rajesh Kumar \textit{et~al.}}

\begin{frontmatter}

\title{Blockchain based Privacy-Preserved Federated Learning for Medical Images: A Case Study of COVID-19 CT Scans}%

\author[1]{Rajesh \snm{Kumar}}
\ead{rajakumarlohano@gmail.com}
\cortext[cor1]{Corresponding author: Wenyong Wang
  Tel.: +86-139-0801-4292;  
}


\author[2]{WenYong  \snm{Wang} \corref{cor1}}
\ead{wywang@must.edu.mo}

\author[4]{Cheng  \snm{Yuan}}
\ead{chy@my.swjtu.edu.cn}


\author[1]{Jay  \snm{Kumar}}
\ead{jay_tharwani1992@yahoo.com}

\author[3]{  \snm{Zakria}}
\ead{zakria.uestc@hotmail.com }


\author[6]{Chengyu  \snm{Zheng}}
\ead{18081000808@189.cn}



\author[3]{Abdullah Aman \snm{Khan}}
\ead{abdkhan@hotmail.com}


\address[1]{Yangtze Delta Region Institute (Huzhou), University of Electronic Science and Technology of China, Huzhou 313001, China.}

\address[2]{International Institute of Next Generation Internet, Macau University of Science and Technology, Taipa, 999078, Macau}

\address[3]{School of Software Engineering, University of Electronic Science and Technology of China, Chengdu, 611731, China.}

\address[4]{Materials Science and Engineering, Southwest Jiaotong University, Chengdu, 611731, China.}
\address[6]{China Telecommunications Corporation,  Sichuan Branch , Chengdu 611731, China}

\received{1 May 2021}
\finalform{10 May 2021}
\accepted{13 May 2021}
\availableonline{15 May 2021}
\communicated{S. Sarkar}

\begin{abstract}
Medical health care centers are envisioned as a promising paradigm to handle the massive volume of data of COVID-19 patients using artificial intelligence (AI). 
Traditionally, AI techniques often require centralized data collection and training the model in a single organization, which is most common weakness due to the privacy and security of raw data communication.  To solve this challenging task, we propose a blockchain-based federated learning framework that provides collaborative data training solutions by coordinating multiple hospitals to train and share encrypted federated models without leakage of data privacy. The blockchain ledger technology provides the decentralization of federated learning model without any central server. The proposed homomorphic encryption scheme encrypts and decrypts the gradients of model to preserve the privacy. More precisely, the proposed framework: i) train the local model by a novel capsule network to segmentation and classify COVID-19 images, ii) then use  the homomorphic encryption scheme to secure the local model that encrypts and decrypts the gradients, and finally the model is shared over a decentralized platform through proposed blockchain-based federated learning algorithm. The
integration of blockchain and federated learning  leads to a new paradigm for  medical image data sharing in the decentralized network. The conducted experimental results
demonstrate the performance of the proposed scheme.

\end{abstract}


\end{frontmatter}



\section{INTRODUCTION}

The drastic spread of novel coronavirus (COVID-19) around the globe has caused a large number of deaths in a year. The COVID-19 virus causes acute respiratory disease, which directly infects the human lungs, resulting in intensive breathing difficulty. Due to the highly contagious nature, COVID-19 detection remains among high-priority tasks. Currently, various artificial intelligence (AI) techniques are under exploration to discover better solutions to detect it  \cite{kumar2021blockchain,kumar2021integration, deng2021trends,khan2020h3dnn}. 
Particularly, a significant portion of the research is focused on CT scan images in the past year as it proved to be a more reliable source to detect the infection. However, these techniques often require a large amount of data from a single source (hospital or research center) to train the classification model to predict more accurately. In contrast, data from a single source lacks the feature distribution variance. TThe less variation in data directly leads to sampling error and high model loss, affecting the diagnosis results in terms of accuracy. The data variation problem can be solved if many hospitals can share the data. However, the reason data confidentiality and privacy restrict multiple hospitals to share the data to train the model. Due to this issue, traditional learning, where only local data is considered, may not fit properly. In contrast, the transfer learning enables sharing the model instead of sharing the data. The transfer learning exploit a general pre-trained model and modifies it with accordance to local data \cite{das2020automated,pathak2020deep, deng2020visual}.  Yet, the sensitivity of a local model totally depends on the quality of the pre-trained model. Let us take a scenario in a rural area hospital with insufficient data to train the model. However, the hospital can collaborate with another hospital while considering the same goal without sharing the data. However, transfer learning still confines the base model to increase more robustness while taking benefit from local data of the hospital. Due to this reason, hospitals are unable to get the full benefit from AI. 

Recently, the federated learning technique is introduced to solve the problem by collaboratively training the model without physically exchanging the model itself. The collaborative model solves the data variance issue and enables the evolution of the model over time for all hospitals. Generally, it is a collaborative learning framework of federated learning which enables multiple collaborators to train their local model and send the learned weights to a centralized server where they are aggregated into a global model.  This procedure of gaining knowledge is in the form of a consensus model without moving patient data beyond the firewalls of parent data recording centers (hospital or research center). To this point, the learning process occurs locally at each participating institution, and only the model characteristics are transferred to a federated server for global model training. Originally federated learning was developed for different domains, such as distributed learning, edge device, and mobile computing \cite{01DBLP:journals/tist/YangLCT19,02DBLP:conf/amcis/ThomasAL18}. Due to its vast scope of applicability, it has gained considerable research attention for healthcare applications \cite{02DBLP:journals/fgcs/BlanquerBBCCFFG20,03DBLP:journals/mia/0005XLMRH0TTWZC21,04DBLP:conf/bigcomp/ThwalTTH21,05DBLP:journals/corr/abs-2101-11693,06DBLP:journals/corr/abs-2101-11693,07DBLP:journals/mia/LiGDSVD20,08DBLP:conf/visapp/BahetiSAR20,09DBLP:journals/corr/abs-2005-11756,10DBLP:journals/jbi/HuangSQMDL19,11DBLP:journals/ijmi/BrisimiCMOPS18,12DBLP:journals/toit/CanE21}. 
Recent research has proven that models trained by federated learning can achieve a comparable levels of performance to ones trained on centrally hosted medical data \cite{12DBLP:journals/toit/CanE21,13DBLP:journals/ton/DinhTNHBZG21,14DBLP:journals/cacm/ChengLCY20,15DBLP:conf/acmturc/Yang0LY20}. However, still there exist security issue of federated learning \cite{shokri2015privacy}, where the users can share the gradients to verify the security and privacy of the data. To this end, their methodology was exposed to vulnerability even for passive attackers \cite{Dai2019,tang2018enabling}.  

To tackle the security and scalability issues, the blockchain as a ledger technology is attractive to provide model decentralization without involving any central server. Particularly blockchain provides the facility to collect the  data model securely from the different points or locations  (i.e., Japan, China, Pakistan, USA, UK) to train the global model. The recent works focus on federated learning using central server topology \cite{02DBLP:journals/fgcs/BlanquerBBCCFFG20,03DBLP:journals/mia/0005XLMRH0TTWZC21,04DBLP:conf/bigcomp/ThwalTTH21,05DBLP:journals/corr/abs-2101-11693,06DBLP:journals/corr/abs-2101-11693,07DBLP:journals/mia/LiGDSVD20,08DBLP:conf/visapp/BahetiSAR20,09DBLP:journals/corr/abs-2005-11756,10DBLP:journals/jbi/HuangSQMDL19,11DBLP:journals/ijmi/BrisimiCMOPS18,12DBLP:journals/toit/CanE21}.  However, none of the above considers blockchain-based federated learning to decentralize the global model in medical image analysis. Therefore, there still exist the gap between secure federated learning without the dependency of a central server to provide secure model sharing and trust issue for data providers (e.g., hospital). 

Motivated by a current situation where the model needs to update continuously to deal with new types of COVID-virus over time while considering the above-discussed issues. In this paper, we propose a framework that integrates privacy-preserving federated learning over the decentralized blockchain. To train the local model, we design a capsule network based on segmentation and classification to detect the COVID-19 images. Our segmentation network aims to extract nodules from the chest CT images. For each locally trained model, gradients are encrypted using a homomorphic encryption technique to preserve the privacy of each hospital. In this encryption technique, the hospitals are assigned the same secret key to reducing the communication overhead for high-dimensional data in neural networks. In this way, the client's or users' side encryption knowledge, which guarantees user privacy and blockchain, ensures the data's reliability. The task of aggregation and learning the global model is trained over the blockchain. We exploit the Direct Acyclic Graph (DAG) to reduce the computation efficiency of the blockchain. The main contributions of this paper are summarized as follows:

\begin{enumerate}
	\item We design blockchain based federated-learning  framework  which provides collaborative data training  and  decentralization of federated learning model without any central server.
	\item We designed a homomorphic encryption scheme for encrypting the weights
	of the local model, which can ensure the hospital data's privacy.
	\item We design a blockchain based federated learning algorithm  to build data models and sharing the data models instead of raw data. It aggregate the local model weights and train the global model.
	\item For local model training, we  propose an Capsule Network model  for segment
	pneumonia infection regions and  automatically classify the COVID-19 chest CT images. 
	\item The proposed framework  update model continuously  to deal with new types of COVID-virus and  easily  share the latest information of the patients through out the world.
\end{enumerate}
The rest of the paper is arranged as follows. Section \ref{sec:PRELIMINARIES}
discuss and introduce the basic knowledge of the deep learning and
blockchain technology. In Section \ref{sec:Blockchain-and-FEDERATED}
, discuss the system model, COVID-19 CT image detection framework,
then design a protocol for encryption gradients. Finally, blockchain
based federated learning model. Then , we discuss the performance
analysis and security analysis in Section \ref{sec:SECURITY-ANALYSIS-AND}.
Finally, we concludes this work in Section \ref{sec:Conclusion}.

\section{PRELIMINARIES}

\label{sec:PRELIMINARIES}

This section briefly introduces the fundamentals of deep learning,
federated learning, homomorphic encryption, and blockchain-based federated
learning model, which is followed by the system model. The main mathematical
notations used in this article are listed in Table \ref{Tab:notations}.

\begin{table}[h]
	\caption{Summary of the notations}
	\label{Tab:notations}
	
	\begin{tabular}{|c|c|}
		\hline 
		Notations & Description\tabularnewline
		\hline 
		\hline 
		$W_{i}(a)$ & Local model weights\tabularnewline
		\hline 
		$m_{i}(t)$ & local model learned by devices\tabularnewline
		\hline 
		$CW$ & The cumulative weight of tr\tabularnewline
		\hline 
		$W$ & weights of the model\tabularnewline
		\hline 
		$P_{x,y}$ & The transition probability of transactions \tabularnewline
		\hline 
		$\lambda$ & The 0 an 1 selection state\tabularnewline
		\hline 
		$\mathbf{}\stackrel{}{\mathbb{Z_{N}}}$ & Plaintext space\tabularnewline
		\hline 
		$\mathbf{A}\stackrel{\$}{\longrightarrow}\mathbb{Z}_{p}^{\kappa\times\tau}$ & Matrix \tabularnewline
		\hline 
		$g$ & Gradients vector for the model\tabularnewline
		\hline 
		$pk/sk$ & Public/Private key \tabularnewline
		\hline 
		$\otimes$ & Product between two ciphertexts\tabularnewline
		\hline 
	\end{tabular}
\end{table}

\subsection{Deep Learning}

\begin{figure}
	\centering
	\includegraphics[scale=0.7]{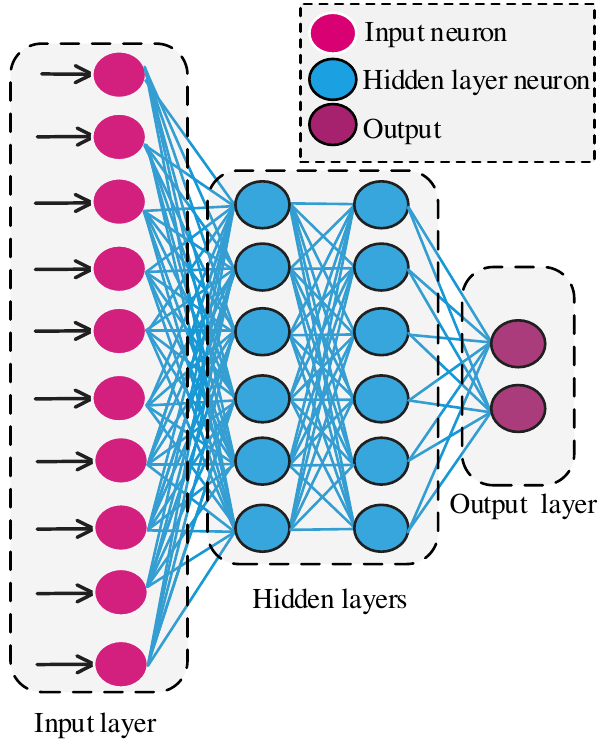}\caption{Background of Basic Deep Learning Model}
	\label{Fig:Deep-Learning}
\end{figure}

The deep learning models are used the feedforward and backpropagation
algorithms to train the model shown in Figure \ref{Fig:Deep-Learning}.
The feedforward function defined as $f(x,w)=y\lyxmathsym{¯}$ , where
$x$ shows the input vector and $w$ represents the parameter vector.
The $D={(x_{i},y_{i});i\in I}$ is the training dataset for the each
instance of $(x_{i},y_{i})$ . Moreover $l$ is the loss function
, whereas the training dataset $D$ on loss function defined as $L(D,w)$
$=\frac{1}{|D|}\sum_{\left(\mathbf{x}_{i},\mathbf{y}_{i}\right)\in D}l\left(\mathbf{y}_{i},f\left(\mathbf{x}_{i},\mathbf{w}\right)\right)$.
However, the backpropagation phase utilized the stochastic gradient
descent (SGD) for updating the parameters.

\begin{equation}
	\mathbf{w}^{t+1}\leftarrow\mathbf{w}^{t}-\eta\nabla_{\mathbf{w}}L\left(D^{t},\mathbf{w}^{t}\right)\label{eq:1}
\end{equation}

where $\eta$ learning rate of the hyperparameter and $w^{t}$ defined
as vector of $i_{th}$iteration. However, $D^{t}$ is the training
dataset. Equation \ref{eq:1} shows the standard training procedure
to train the data for the one hospitals or users.

\subsection{Federated Learning}

Federated learning is a distributed and secure deep learning technique
that enables training a shared model without leakage the hospitals
privacy. Moreover, federated learning has introduced a mechanism to
collect the data from various parties or hospitals without leakage
the hospitals privacy. The advantage of the federated learning model
is reducing the resources (i.e., memory, power) consummation of a
single participant and improving the quality of the training model.
In other words, federated learning learn the model collaboratively
and share the trained model in the local machines. More detail, each
users $u\in U$ has own private dataset $D_{u}\subseteq D$. The equation
for the mini-batch dataset $D^{t}=\bigcup_{u\in U}D_{u}^{t}$ with
SGD shown below 

\begin{equation}
	\mathbf{w}^{t+1}\leftarrow\mathbf{w}^{t}-\eta\frac{\sum_{u\in u}\nabla_{\mathbf{w}}L\left(D_{u}^{t},\mathbf{w}^{t}\right)}{|U|}\label{eq:2}
\end{equation}

\begin{figure}[h]
	\centering
	\includegraphics[scale=0.3]{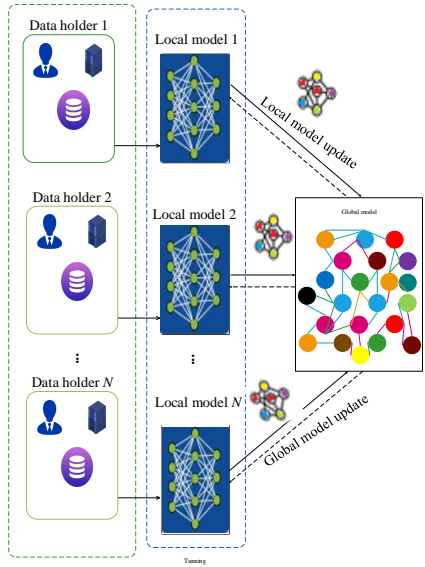}\caption{Federated Learning Model}
	\label{Fig:federated-learning}
\end{figure}

Each user shares the local model to the blockchain distributed ledger
for training the global shared model. Then, the users / hospitals
upload the new data, i.e., (gradients or weights) for updating the
global model. Moreover, each user $u\in U$ has own private dataset
with data samples for federated learning which is shown in Figure
\ref{Fig:federated-learning}. 

\begin{equation}
	F_{i}(w)=\frac{1}{\left|D_{i}\right|}\sum_{j\in D_{i}}f_{j}\left(w,x_{i},y_{i}\right)\label{eq:3}
\end{equation}

For multiple devices or hospitals with dataset $D$, a global loss
\cite{zhu2019multi} function $f_{j}\left(w,x_{i},y_{i}\right)$ minimizing
the weights. The difference between the estimated and real for each
hospital $f_{j}\left(w,x_{i},y_{i}\right)$ , the global model function
of the $F(w)$ is described as 

\begin{equation}
	F(w)=\frac{1}{\left|M_{I}\right|}\sum_{i\in I}u_{i}\cdot F_{i}(w)=\frac{1}{\left|M_{I}\right|}\sum_{i\in I}\sum_{i\in D_{i}}u_{i}\cdot\frac{f_{j}\left(w,x_{i},y_{i}\right)}{\left|D_{i}\right|}\label{eq:4}
\end{equation}

Where $i$ is sample dataset $(x_{i},y_{i})$ of the gallery $I=\{1,2,\cdots,n\}$
\cite{tran2019federated} , $u_{i}$ is the number of hospitals individual
dataset models. In our proposed training process, we enhanced the
accuracy of the model by iteratively minimizing the loss function
of the global model. The equation of the loss function given as 

\begin{equation}
	Q(w,t)=\underset{i\in I,t\leq T}{\arg\min}F(w)\label{eq:5}
\end{equation}

\begin{equation}
	Pr\left(w_{i}\in\mathbb{R}_{d}\right)\leq\exp(\epsilon)Pr\left(w_{i}^{\prime}\in\mathbb{R}_{d}\right)\label{eq:6}
\end{equation}

\begin{equation}
	\sum_{i=1}^{t}\Delta t(i)\leq\min\left(T_{1},T_{2},\ldots,T_{n}\right)\label{eq:7}
\end{equation}

Where $Pr\left(w_{i}\in\mathbb{R}_{d}\right)\leq\exp(\epsilon)Pr\left(w_{i}^{\prime}\in\mathbb{R}_{d}\right)$
is the privacy of the users \cite{lu2019differentially} of the parameters
of the $\left(T_{1},T_{2},\ldots,T_{n}\right)$. $\Delta t(i)$ is
the execution time of the iteration.

\subsection{Homomorphic Encryption }

Homomorphic encryption allows the calculation of encrypted data (ciphertext)
without decryption. The new encrypted data matches the result of the
operation performed on the unencrypted data after decryption. We utilized
the BGV \cite{brakerski2014leveled} ncryption scheme, which takes
as input the secret key with large noise and outputs an unencrypted
data of the same data with a fixed amount of noise. Additionally,
a key-switching procedure which text encrypt data and output the same
message. We refer to the detailed encryption scheme for readers in
\cite{brakerski2014leveled}. Therefore, we used homomorphic encryption
to encrypt the gradients \cite{aono2017privacy,bottou2010large} to
share the data in the blockchain distributed network. The previous
research shares the encrypted gradients and shares to the centralized
server\cite{li2015personalized,li2014enabling}. They do not consider
a distributed blockchain network. The problem of a blockchain database
is cost-effective. For that reason, we use homomorphic to encrypt
the model and train the local model to the global model.

We define $Z$ is the unencrypted matrix data of the mini-batch dataset
with the size of $S*T$, before the encryption of tensor, a private
key matrix $\phi$ with size $S*S$ as: 

\begin{equation}
	\left[\begin{array}{cccc}
		\phi_{11} & \phi_{12} & \cdots & \phi_{1S}\\
		\phi_{21} & \phi_{22} & \cdots & \phi_{2S}\\
		\vdots & \vdots & \vdots & \vdots\\
		\phi_{S1} & \phi_{S2} & \cdots & \phi_{SS}
	\end{array}\right]\label{eq:8}
\end{equation}

This key only access by the users/participants who share the mini-batch
dataset 

\begin{equation}
	\left[\begin{array}{c}
		\mathbb{Z}_{(1)}\\
		\mathbb{Z}_{(2)}\\
		\vdots\\
		\mathbb{Z}_{(S)}
	\end{array}\right]=\left[\begin{array}{cccc}
		\phi_{11} & \phi_{12} & \cdots & \phi_{1S}\\
		\phi_{21} & \phi_{22} & \cdots & \phi_{2S}\\
		\vdots & \vdots & \vdots & \vdots\\
		\phi_{S1} & \phi_{SS2} & \cdots & \phi_{SS}
	\end{array}\right]\otimes\left[\begin{array}{c}
		Z_{(1)}\\
		Z_{(2)}\\
		\vdots\\
		Z_{(N)}
	\end{array}\right]\label{eq:9}
\end{equation}

where $Z(i)$ shows the vector data of the $i_{th}$ node of the blockchain
ledger. The $\otimes$ operator shows the product between two ciphertext

\begin{equation}
	\mathbb{Z}_{(i)}=\phi_{i1}Z_{(1)}+\phi_{i2}Z_{(2)}+\cdots+\phi_{iN}Z_{(S)}\label{eq:10}
\end{equation}

\begin{figure}
	\includegraphics{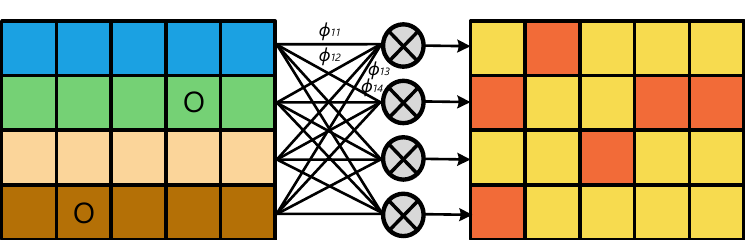}\caption{Homomorphic encryption}
	
	\label{fig:Homomorphic-encryption}
	
\end{figure}

The Figure \ref{fig:Homomorphic-encryption} shows the homomorphic
encryption function with the linear transformation of matrix. In this
way, the linear transformation maintain the low rank functionality.
The function $\phi_{ij}\in[0,1)$, and $\sum_{j=1}\psi_{i,j}=1$ shows
the homomorphic encryption with private key. 

\subsection{Blockchain-Enabled Federated Learning}

To train the better AI model for the industry 4.0 required to collect
the data from multiple sources without leakage the privacy and authentication
of the users. Therefore, we use federated learning with the blockchain
distributed ledger to update the global AI model. The blockchain collects
the data model from different nodes and aggregates the local and global
model. The smart contract uploaded the weights and updated the models.
The proposed architecture integrates blockchain with federated learning
for full decentralization and enhancing security. Also, decentralization
provides more accuracy of the model and enables the poisoning-attack-proof.

Some issues are not resolved for federated learning, i.e., insufficient
incentives, poisoning attacks, etc. Therefore some authors\cite{Lu2020,qu2020decentralized}
design the blockchain with the federated learning. Similarly, Pokhrel
and Choi \cite{pokhrel2020federated} design a technique to protect
privacy. The major issue of previous papers not including the encryption
technique with the blockchain model gradients sharing. Therefore,
this paper use the directed acyclic graph with the with the Proof-of-Work
(PoW) consensus algorithm for the aggregation of gradients. Additionally,
this work is fully decentralized and train accurate model without
leakage the privacy of the user.

\section{SECURE DATA SHARING FOR BLOCKCHAIN AND FEDERATED LEARNING }

\label{sec:Blockchain-and-FEDERATED}

\begin{figure*}
	\centering
	\includegraphics[height=10cm,width=18cm]{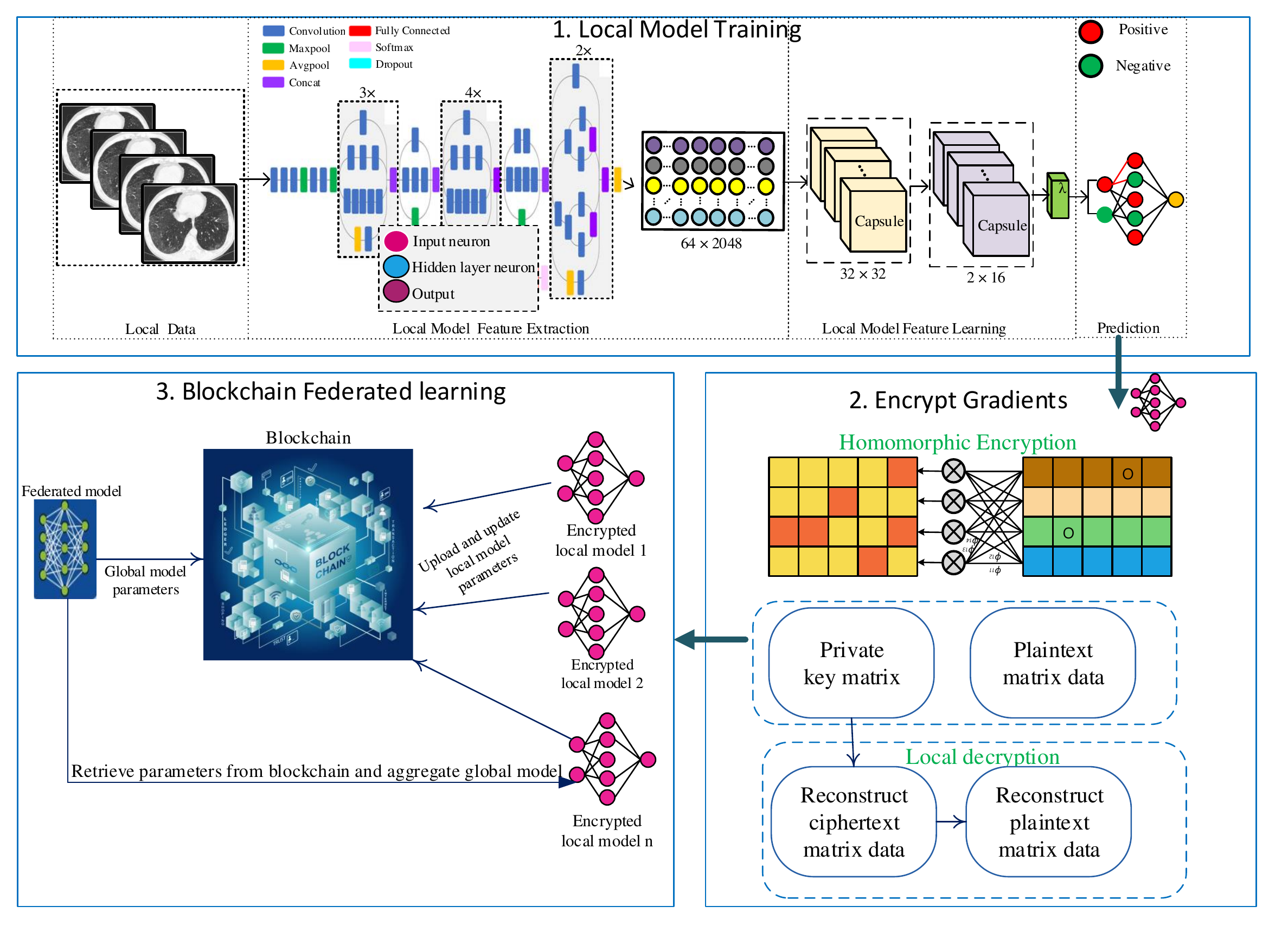}\caption{Local Model: Deep Learning Model for COVID-19. We employ a modified
		version of inception V3 (IV3{*})deep learning model as a feature extraction
		pipeline. Further, we train the extracted features using to layers
		of the capsule network. Encypt Gradiants: encrypt the weights. Blockcahin
		federated learning combine the local model.}
	\label{Fig:1}
	
\end{figure*}

In this section, We first we introduce the high level architecture
of the system and technical details in Figure \ref{Fig:1}. Our proposed
scheme consist of multiple users share the data securely using the
federated learning with blockchain technology. The proposed architecture
has multiple phases. 

\textbf{Local model:} 
\begin{enumerate}
	\item Input COVID-19 images to train the model.
	\item Learn the local model and calculate the local gradients.
	\item Encrypt the gradients of the local model.
\end{enumerate}
\textbf{Send the weights to the blockchain network for aggregation
	model: }
\begin{enumerate}
	\item Aggregate $W_{i}(a)\leftarrow\frac{1}{\sum_{i\in\mathcal{S}_{t}}\left|\mathcal{D}_{i}\right|}\sum_{i\in\mathcal{S}_{t}}\left|\mathcal{D}_{i}\right|W_{i}(a)$
	all users weights ciphertext. 
\end{enumerate}
\textbf{Broadcast the weights: }
\begin{enumerate}
	\item Update the deep learning model based on global weights.
	\item Upload the local model updates.
\end{enumerate}

\subsection{Training The Local Model For the COVID-19 dataset}

In this section, we train the local model for detection of the COVID-19.
The main model divided into three parts: (i) Segmentation Network
(ii) Classification (iv) Probabilistic Grad-CAM Saliency Map Visualization

\subsubsection{Segmentation network}

Our segmentation network obtains the ground-truth lung masks and extracts
the lung region using a learning method \cite{liao2019evaluate,lalonde2018capsules}.
We removed the unnecessary or failure data manually, and the renaming
segmentation data was taken as ground-truth masks. The 3D Lung mask
is input the whole image for training and testing data. The training
objective is to adopt the capsule network segmentation. Where $r_{t_{i}^{\ell}\mid xy}$
is the routing coefficient, $b_{t_{i}^{\ell}\mid xy}$ shows the pixel
of images, s and y shows the ground truth label with the $\in$ \{heart,
background, left lung, right lung\}. 

\begin{equation}
	r_{t_{i}^{\ell}\mid xy}=\frac{\exp\left(b_{t_{i}^{\ell}\mid xy}\right)}{\sum_{k}\exp\left(b_{t_{i}^{\ell}k}\right)}\label{eq:10-1}
\end{equation}

To determine the final output of the segmentation using the non-linear
squashing function 

\begin{equation}
	\mathbf{v}_{xy}=\frac{\left\Vert \boldsymbol{p}_{xy}\right\Vert ^{2}}{1+\left\Vert \boldsymbol{p}_{xy}\right\Vert ^{2}}\frac{\boldsymbol{p}_{xy}}{\left\Vert \boldsymbol{p}_{xy}\right\Vert }\label{eq:11}
\end{equation}

Where $\mathbf{v}_{xy}$ is the output of the segmented image with
the spatial location $(x,y)$ and $\boldsymbol{p}_{xy}$ is the final
input. 

\subsubsection{Classification}

 We design a Capsule Network because it achieves high performance in detecting
diseases in the medical images. The previous technique needs lots
of data to train a more accurate model. The Capsule Network improves
the deep learning models' performance inside the internal layers of
the deep learning models. The architecture of our modified Capsule
Network is similar to Hinton Capsule Network. The Capsule network
contains four layers: i)convolutional layer, ii) hidden layer, iii)
PrimaryCaps layer, and iv) DigitCaps layer.

A capsule is created when input features are in the lower layer. Each
layer of the Capsule Network contains many capsules. To train the
capsule network, the activation layer represents instantiate parameters
of the entity and compute the length of the capsule network to re-compute
the scores for the feature part. Capsule Networks is a better replacement
for Artificial Neural Network (ANN). Here, the capsule acts as a neuron.
Unlike ANN where a neuron outputs a scalar value, capsule networks
tend to describe an image at a component level and associate a vector
with each component. The probability of the existence of a component
is represented by this vectors length and replaces max-pooling with
"routing by agreement". As capsules are independents the probability
of correct classification increases when multiple capsules agree on
the same parameters. Every component can be represented by a pose
vector $U_{i}$ rotated and translated by a weighted matrix $W_{i,j}$
to a vector $\hat{u}_{i|j}$. Moreover, the prediction vector can
be calculated as:

\begin{equation}
	\hat{u}_{i|j}=W_{i,j}u_{i}
\end{equation}
The next higher level capsule i.e., $s_{j}$ processes the sum of
predictions from all the lower level capsules with $c_{i,j}$ as a
coupling coefficient. Capsules $s_{j}$ can be represented as: 
\begin{equation}
	S_{j}=\sum_{i}c_{i,j}\hat{u}_{i|j}
\end{equation}
where $c_{i,j}$ can be represented as a routing softmax function
given as: 
\begin{equation}
	c_{i,j}=\frac{e^{b_{ij}}}{\sum_{k}e^{b_{ik}}}
\end{equation}
As can be seen from the Figure \ref{Fig:1}, the parameter c, A squashing
function is applied to scale the output probabilities between 0 and
1 which can be represented as: 
\begin{equation}
	a=\frac{\|a\|^{2}}{1+\|a\|^{2}}\frac{a}{\|a\|}
\end{equation}

For further details, refer to the original study \cite{DBLP:conf/nips/SabourFH17}.

\subsubsection{CAM map visualization}

We find the interpretability of the proposed capsule network by visualization
of the COVID-19 slices. The most widely (GRAD-CAM) technique previous
technique \cite{selvaraju2017grad}. More precisely,The GRAD-CAM map
takes input as an image using the following equation. 

\begin{equation}
	l^{c}(x)=Upsampling\left(\sigma\left(\sum_{M}\alpha_{M}^{c}f^{M}(x)\right)\right)\in I\label{eq:15}
\end{equation}

where $I$ is the input image is the last layer of the convolution
layer. Moreover, upsampling of the input image $m*n$ with the feature
vector$u*v$. $\sigma$ defined as the ReLU layer. However the probability
of each pixel calculated by 

\begin{equation}
	\left[l_{prob}^{c}(x)\right]_{i}=\frac{1}{M_{i}}\left[\sum_{M=1}^{M}r^{c}\left(x_{M}\right)\mathcal{Q}_{M}\left(l^{c}\left(x_{M}\right)\right)\right]_{i}\label{eq:16}
\end{equation}

Where $K$ is the slice of the each image $x$ pixel, $\boldsymbol{l}^{c}\left(\boldsymbol{x}_{M}\right)$
compute the GRAD-CAM by using the equation \ref{eq:15} with respect
to frequency of the image. $M$ is computed after the soft max layer
of the capsule network. Equation \ref{eq:16} shows the average probability
of the each pixel of the class for the global saliency map.

\subsection{The Architecture of Gradients Encryption \& Decryption }

The data provider P, who holds the private medical images $I$, rain
the local model and encrypt the local model vector. Then send to the
blockchain network B. The blockchain federated learning model aggregates
the encrypted vector using the global federated learning model. Moreover,
the gradients encryption \& decryption techniques for the secure weight
sharing proposed by Lyubashevsky et al. \cite{elgamal1985public}
based on Ring-LWE scheme. Suppose $\Phi_{n}(X)$ is the reducible
polynomial function The degree of the polynomial function $\phi(n)$
, $R_{p}=R/pR$ and $R=\mathbb{Z}[X]/\left(\Phi_{n}(X)\right)$ is
the polynomial ring . The samples $(a,b=s\cdot a+e)$ of the Ring-LWE
, where $s,e$ indicates the Gaussian distribution.

Firstly, we define a ciphertext and plaintext space. Ring $R_{p}=(\mathbb{Z}/q\mathbb{Z})[X]/\left(\Phi_{n}(X)\right)$
defined as plaintext with the modulus $q$. Similarity, $\left.R_{p_{1}}=\left(\mathbb{Z}/p_{1}\mathbb{Z}\right)\mid X\right\rceil /\left(\Phi_{n}(X)\right)$
defined as internal ciphertext RBGV and $R_{p_{1}}^{\prime}=\left(\mathbb{Z}/p_{1}\mathbb{Z}\right)[X]/\left(\Phi_{n^{\prime}}(X)\right)$
defined as external ciphertext. However, the $\phi(n)=2l\phi(n)$
with the $l=\left\lceil \log p_{1}\right\rceil $ and $p_{1}=p\cdot p_{0}$
for primes $p,p_{0}$.

Then, we describe some widely used sampling subroutines for better
readability as follows:
\begin{enumerate}
	\item $\mathcal{ZV}(n)$: is represents as a vector space of the $n$ numbers
	from $\{-1,0,1\}$ the probabilities of the each element are $Pr_{-1}=\frac{1}{4},Pr_{0}=\frac{1}{2},Pr_{1}=\frac{1}{4}$ 
	\item $GM(n,\sigma)$: is represents as a vector space of the $n$ numbers,
	the Gaussian distribution $\sigma$ and standard deviation mean $0$.
	\item $\mathcal{V}\mathcal{N}(n,p)$: is represents as a vector space of
	the $n$ numbers from randomly uniform distribution modulo $p$.
\end{enumerate}

\subsubsection{Setup}

Suppose $N\in N$ is number of devices, and $K$ is the security parameter,
For more details, the internal encryption defined as:
\begin{enumerate}
	\item $Draw\tilde{a}\leftarrow\mathcal{V}\mathcal{N}\left(\phi(n),p_{1}\right)\text{ and }\widetilde{s},\widetilde{\gamma}\leftarrow\mathcal{G}M(\phi(n),\tilde{\sigma})$.
	\item $\text{ Compute }b=\widetilde{a}\cdot\widetilde{s}+q\cdot\widetilde{\gamma}$
	\item Output $pk=(a,b)\in R_{p_{1}}\times R_{p_{1}}$ as public key and
	$SK_{C_{2}}=\widetilde{s}\in R_{p_{1}}$ as part of secret key for
	the distributed ledger blockchain.
	\item Output $sk_{i}=s_{i}\in R_{p_{1}}^{\prime}$ as secret key for participant
	$i$ and $SK_{C_{1}}=-\sum_{i}s_{i}\in R_{p_{1}}^{\prime}$ as another
	part of secret key for the distributed ledger blockchain.
\end{enumerate}

\subsubsection{Gradients encryption}

In order to connection among the vector $Z^{n}$ and ring $R$ during
encryption phase, the mappings as follows: 
\begin{enumerate}
	\item $Map_{R\rightarrow Z}n:$ A coefficient representation of a input
	ring elements of $n$ entities. 
	\item $Map_{\mathbb{Z}^{n}}\rightarrow R^{:}$ A matrix over the same ring
	as the vector containing the coefficients representations of the vector
\end{enumerate}
Generation of internal ciphertext (e.g., RBGV)raining Local Models 

\subsubsection{The architecture of gradients encryption decryption for external
	ciphertext }
\begin{enumerate}
	\item Set $\mathbf{v}_{i}=Map\left(\widetilde{c}_{i,0}\|\widetilde{c}_{i,1}\right)\in\mathbb{Z}_{p_{1}}^{2\phi(n)}$
	\item Invoke algorithm 1 to sample $\mathbf{e}_{i}\in\mathbb{Z}_{p1}^{2\phi(n)l}$
	subject to the distribution $\Lambda_{\mathbf{v}_{i}}^{\perp}(\mathbf{G}),$
	where $\mathbf{e}_{i}=sample\left(v_{i_{1}},\sigma\right)$....$.....sample\left(v_{i_{2}\phi(n)},\sigma\right)$
	\item Set $e_{i}=\left(Map_{\left.Z^{\phi\left(n^{\prime}\right.}\right)\rightarrow R_{n!}^{\prime}}\left(\mathbf{e}_{i}\right)\right)$
	\item Compute $c_{i}=a\cdot s_{i}+e_{i}\in R_{p1}$
	\item Send final ciphertext $c_{i}$ to the blockchain network. 
	\item Aggregate all the ciphertexts $c=\sum_{i}c_{i}=a\cdot\sum_{i}s_{i}+$$\sum_{i}e_{i}\in R_{p_{i}}^{\prime}$
	in the blockchain network
	\item Compute the sum of errors terms \$$e=c+a\cdot SK_{C_{1}}=$ $\sum_{i}e_{i}\in R_{p_{1}}^{\prime},$where
	$SK_{C_{1}}=-\sum_{i}s_{i}$.
	\item Set $\text{ }\mathbf{e}=Map_{R_{p1}^{\prime}\rightarrow\mathbb{Z}^{\phi\left(n^{\prime}\right)}}(e)$
\end{enumerate}

\subsubsection{The architecture of gradients encryption decryption for internal
	ciphertext }
\begin{enumerate}
	\item Set $gd_{i}^{t}=Map_{\left.\mathbf{Z}^{\phi(n)}\rightarrow R_{q}^{(}\mathbf{g}\mathbf{d}_{i}^{t}\right)}\in R_{q}$.
	\item Draw $e_{0},e_{1}\leftarrow\mathcal{GS}(\phi(n),\sigma)\text{ and }v\leftarrow\mathcal{ZV}(\phi(n))$.
	\item Compute $\widetilde{c}_{i,0}=\widetilde{b}\cdot\tilde{v}+q\cdot\widetilde{e}_{0}+gd_{i}^{t}\text{ and }\widetilde{c}_{i,1}=\widetilde{a}\cdot\widetilde{v}+q$
	$e_{1}$ for modulus $p_{1}$.
	\item Output internal ciphertext $\widetilde{c}_{i}=\left(\widetilde{c}_{i,0},\widetilde{c}_{i,1}\right)\in R_{p_{1}}\times R_{p_{1}}$
	\item Recover the sum of RBGV ciphertext by computing $\mathbf{v}=\sum_{i}\mathbf{v}_{i}=\mathbf{G}\cdot\mathbf{e}\bmod p_{1}\in\mathbb{Z}_{p1}^{2\phi(n)}$
	\item Split the vector $\mathbf{v}=\left(\mathbf{c}_{0},\mathbf{c}_{1}\right)\in\mathbb{Z}_{p_{1}}^{\phi(n)}\times\mathbb{Z}_{p_{1}}^{\phi(n)}$.
	\item Set $\tilde{c}_{0}=Map_{\mathbb{Z}^{\phi(n)}\rightarrow R_{p1}}\left(\mathbf{c}_{0}\right)\in R_{p_{1}}$
	and $\tilde{c}=Map_{\mathbf{Z}^{\phi}(n)\rightarrow R_{p1}}\left(\mathbf{c}_{1}\right)\in R_{p_{1}}$
	\item Invoke algorithm Scale $(\left(\widetilde{c}_{o},\widetilde{c}_{1}\right),p_{1},p_{0}$
	) to switch modulus and produce the scaled ciphertext $\tilde{c}_{o},\tilde{c}_{1}$
	modulo $p_{0}$
	\item Decrypt the ciphertext and produce the sum of plaintext by $gd^{t}=\sum_{i\in[N]}gd_{i}^{t}=\widetilde{c}_{o}-SK_{C_{2}}\cdot\vec{c}_{1}\quad\bmod q\in R_{q}$
	\item Set${gd}^{t}=Map_{R_{q}\rightarrow\mathbf{Z}^{\phi(n)}}\left(gd^{t}\right)$.
	\item Broadcast the global gradients $\mathrm{gd}^{t}$
\end{enumerate}

\subsection{Consensus in Permissioned Blockchain Federated Learning }

\label{subsec:Consensus-in-Permissioned}

The main goal of this section is to exaggeration of the global model
with the blockchain DAG mechanism. The local DAG is responsible for
synchronous global training via federated learning. However, the storage
capability is improved to store the model in the DAG. Based on the
federated learning and permissioned blockchain, the following steps
to adjust the decentralized model aggravation. Firstly select the
nodes of the users and then local train and encrypt the weights. Then
aggregate the weights in the global model. The consensus (i.e., POW)
for data sharing is high cost. To address the problem of the high
cost, we proposed a hybrid DAG based scheme is provided in Algorithm
2. However, we combine the update weight process of federated learning
with the quality verification process using the blockchain DAG. The
Algorithm \ref{Alg:GlobalModel} shows the global aggregation of the
model gradients for the federated learning. 

\begin{algorithm}
	\SetAlgoLined 
	
	$\theta_{\text {global }}^{t-1}$  $\gets$ global model\;
	$\left\{G_{I(j)}^{t}\right\}_{j=1}^{m}$  $\gets$  legal gradient vectors\;
	$g^{t}_{global}$  $\gets$ 0\;
	$l$ $\gets$ 0\;
	\For{k=1,2,..m}{
		\If{$G_{I(k)}^{t} \neq \perp$}{
			Compute $G_{\text {global}}^{t} \leftarrow G_{\text {global}}^{t}+\alpha_{I(k)} l_{l(k)} G_{I(k)}^{t} $\;
			Compute $\leftarrow l+\alpha_{I(k)} l_{I(k)}$
		}
		Compute $G_{g l o b a l}^{t} \leftarrow \frac{1}{I} G_{g l o b a l}^{t}$ \;
		update $\theta_{\text {global }}^{t} \leftarrow \theta_{\text {global }}^{t-1}-\eta G_{\text {global }}^{ }$
	}

	\label{Alg:GlobalModel}
	\caption{Global Federated Learning aggregation algorithm. } 
\end{algorithm}

\subsubsection{The local directed acyclic graph (DAG)}

The local DAG structure is used individually for the each user. In
each iteration $t$ represents federated learning, permissioned blockchain
nodes are selected to verify the aggregation of model $u_{a}$. In
local weight aggregation of deep learning model, weights $u_{i}\in u_{P}$
are transfer to the updated model $m_{i}(t)$ to the near by users.
The model accuracy of weights $W(m_{i}(t))$ is calculated as 
\begin{equation}
	W\left(m_{i}(t)\right)=\frac{\left|d_{i}\right|+\rho\cdot\sum_{j}d_{m_{j}}}{\sum_{i=1}^{N}\left|d_{i}\right|+\sum_{j}d_{m_{j}}}\cdot s_{i}\cdot Acc\left(m_{i}(t)\right)\label{eq:17}
\end{equation}
Where i is the local training and $\left|d_{i}\right|$ is the dataset
size of the model, \textbf{$\sum_{j}d_{m_{j}}$ }represents the accumulated
dataset size of the deep learning local model. $S_{i}$ execute the
each user training slots and $Acc\left(m_{i}(t)\right)$ shows the
accuracy of the each trained model.

To verify the reliability of the transaction weights , we calculate
weight transaction $CW(m_{i}(t))$ 

\begin{equation}
	CW\left(m_{i}(t)\right)=W\left(m_{i}(t)\right)+\frac{1}{M}\sum_{j=1}^{M}\Delta Acc_{j}\cdot W(j)\label{eq:18}
\end{equation}

Where $\Delta Acc_{j}=Acc_{j}\left(m_{i}(t)\right)-W\left(m_{i}(t)\right),W(j)$
are the weight of the each transacation j, where $m_{i}(t)$. $Acc_{j}$
verifies the accuracy of the $m_{i}(j)$

\subsubsection{Add the transaction into the blockchain DAG}

To add the transaction to the blockchain DAG to update the deep learning
model, first required to validate the local model's two transaction
accuracy. Then attach all hashes and generate a new block. The new
block (new transaction) is updated the blockchain DAG, which can broadcast
the nodes in the local model blockchain DAG. The Markov-chain Monte
Carlo prototype is used to check the probability of every step. The
equation of the Markov-chain Monte Carlo is defined as : 

\begin{equation}
	\begin{aligned}E[f(x)]\approx & \frac{1}{m}\sum_{i=1}^{m}f\left(x_{i}\right)\\
		& \left(x_{0},x_{1},\ldots,x_{m}\right)\sim MC(p)
	\end{aligned}
	\label{eq:19}
\end{equation}

\subsubsection{Confirmation and consensus}

The transactions are confirmed or validated based on the cumulative
weights. This article utilized the weighted walk method based on credibility,
which can validate the transaction by selecting the unverified transactions.
When a new transaction is generated, two walkers will be added to
the blockchain DAG to select the transaction. The more transaction
has been pass for verification to achieve high cumulative weight for
verification. 

\begin{equation}
	P_{xy}=\frac{e^{CW(y)-CW(x)}}{\sum_{z:z\rightarrow x}e^{CW(z)-CW(x)}}\label{eq:20}
\end{equation}

Where $P_{xy}$ is the transition probability towards the unverified
transaction of $x$ and $y$. $z$ defined as the neighboring node
of transaction that belongs to x, and $y\in\{z:z\rightarrow x\}$ 

In this way, the complexity of the PoW is less than traditional PoW.
The more and more transaction is executed then the DAG will be faster
and safer. 
\begin{algorithm}
	\SetAlgoLined 
	
	$D_1$ $\gets$ $\left\{M_{1}, m_{2}, \ldots, v_{N}\right\}$  dataset \;
	$m_{0}$   $\gets$ Initialize global weights with the permissioned blockchain BC and DAG  \; 
	$r_{0}$  $\gets$  select the users to $M_{P} \subset M_{I}$ by the node selection  $\left\{r_{1}, r_{2}, \ldots, r_{N}\right\}$\; 
	
	\For{e $\in[episode]$}{
		Select the leader $r_{0}$ \;
		\For{t $\in[time slot]$}{
			\For{$ D_{i}$ dataset $\in M_{p}$}{
				$m_{i}$ matrix global model  $M_{t-1}$ from permissioned blockchain BC \;
				$m_{i}$ = local traning $w_{i}(t)=w(t)-\eta \cdot \nabla F_{i}\left(w_{t-1}\right)$\;
				$m_{i}$ = get local models  updates DAG\;
				$m_{i}$ run  the local aggregation model and get the updated local model $mi_{t}$ \;
				$m_{i}$ add the transactions to the DAG \;
				
			}
		}
		$r{0} \leftarrow $$(t)=\frac{\sum_{i=1}^{N} C_{i} w_{i}(t)}{\sum_{i=1}^{N} C_{i}}$ DAG blockchain updated the model , and averaging the models into $M(e)$\;
		
		$r{0}$ broadcasts model $M(e)$ to other nodes for verification, add  all the transacations into the blockchain ledger ;
		$r{0}$ include the  $M(e)$ global model  form the blockchain ledger\;
		
	}

	\caption{Federated Learning  Empowered  with Blockchain Network} 
\end{algorithm}


\section{SECURITY ANALYSIS AND PERFORMANCE ANALYSIS}

\label{sec:SECURITY-ANALYSIS-AND}

\subsection{Dataset}

In the past, Artificial intelligence (AI) has gained a reputable position
in the field of clinical medicine. And in such chaotic situations,
AI can help the medical practitioners to validate the disease detection
process, hence increasing the reliability of the diagnosis methods
and save precious human lives. Currently, the biggest challenge faced
by AI-based methods is the availability of relevant data. AI cannot
progress without the availability of abundant and relevant data.

\begin{figure}[h]
	\centering \includegraphics[width=0.99\columnwidth]{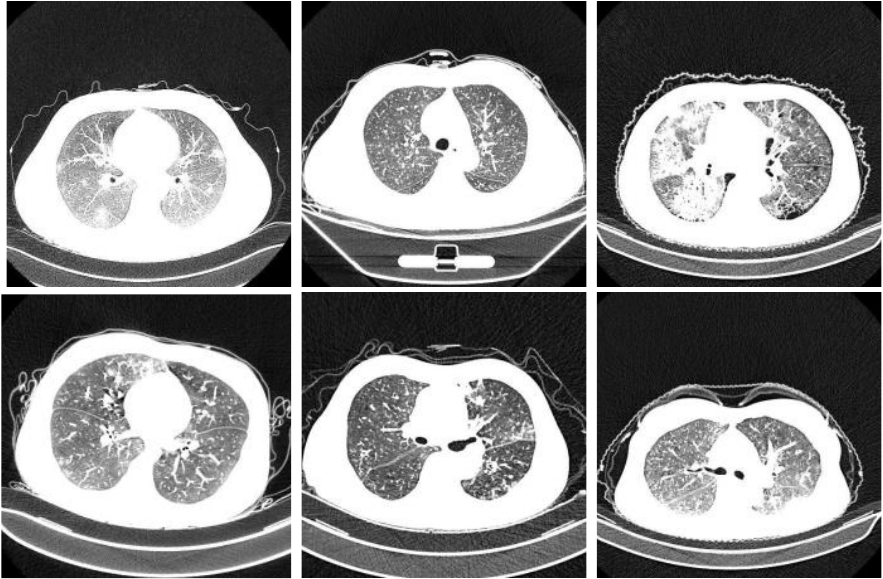}
	\caption{Some random samples of CT scan 2D slices taken from CC-19 dataset.}
	\label{fig:2dssampels}
\end{figure}

\begin{figure*}
	\centering \includegraphics[scale=0.7]{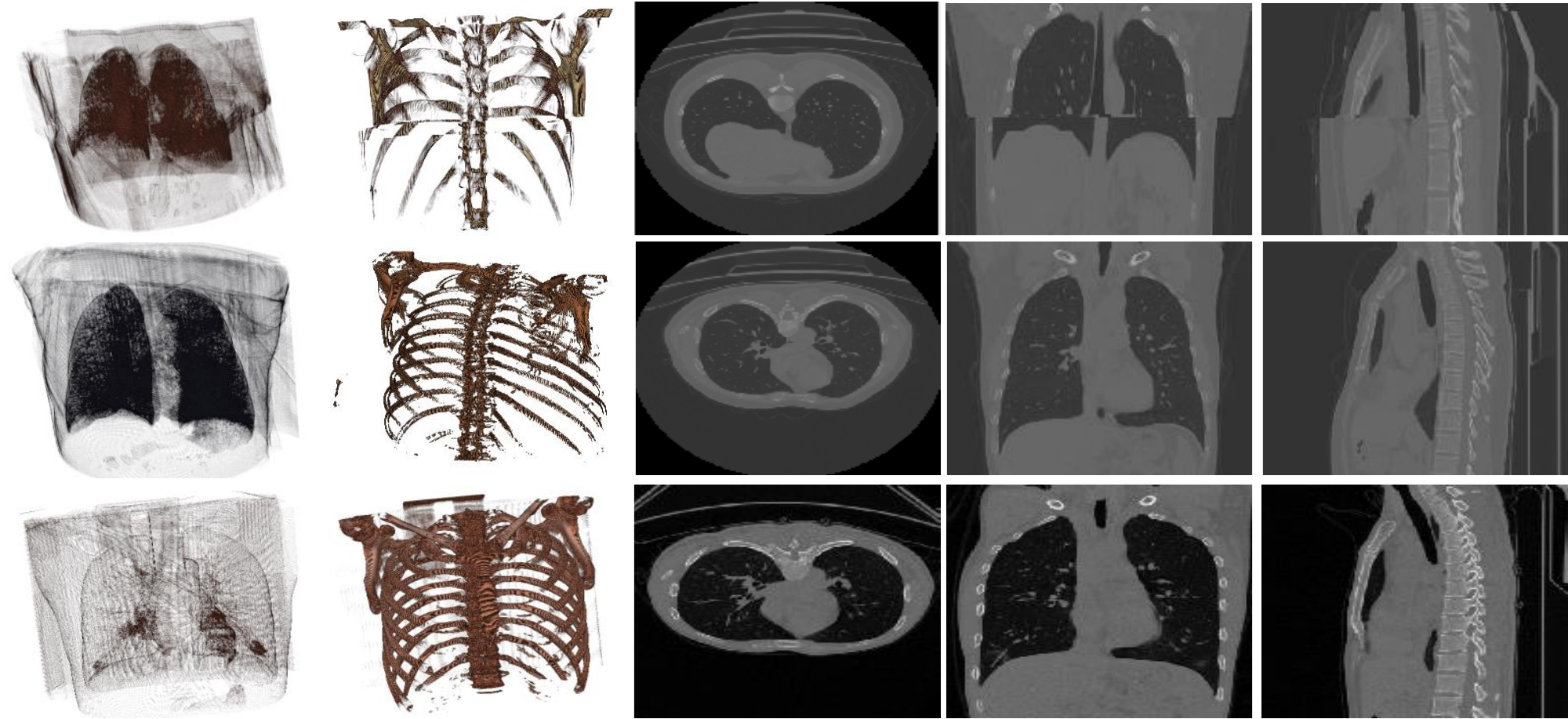} \caption{This figure shows some selected samples from the ``CC-19 dataset".
		Each row represents different patient samples with various Hounsfield
		Unit (HU) for CT scans. The first column, from left to right, shows
		the lungs in the 3D volume metric CT scan sphere. The second column
		shows the extracted bone structure using various HU values followed
		by the XY, XZ, and YZ plane view of the subjects' CT scan. It is worth
		noting that the 3D volumetric representation is not pre-processed
		to remove noise and redundant information.}
	\label{fig:3dssampels}
\end{figure*}

In this paper, we introduce a small new dataset related to the latest
family of coronavirus i.e. COVID-19. Such datasets play an important
role in the domain of artificial intelligence for clinical medicine
related applications. This data set contains the Computed Tomography
scan (CT) slices for 89 subjects. Out of these 89 subjects, 68 were
confirmed patients (positive cases) of the COVID-19 virus, and the
rest 21 were found to be negative cases. The proposed dataset CC-19
contains 34,006 CT scan slices (images) belonging to 89 subjects out
of which 28,395 CT scan slices belong to positive COVID-19 patients.
This dataset is made publicly available via GitHub (https://github.com/abdkhanstd/COVID-19).
Figure \ref{fig:2dssampels} shows some 2D slices taken from CT scans
of the CC-19 dataset. Moreover, some selected 3D samples from the
dataset are shown in Figure \ref{fig:3dssampels}. The Hounsfield
unit (HU) is the measurement of CT scans radiodensity as shown in
Table \ref{tab:hu}. Usually, CT scanning devices are carefully calibrated
to measure the HU units. This unit can be employed to extract the
relevant information in CT Scan slices. The CT scan slices have cylindrical
scanning bounds. For unknown reasons, the pixel information that lies
outside this cylindrical bound was automatically discarded by the
CT scanner system. But fortunately, this discarding of outer pixels
eliminates some steps for preprocessing.

\begin{table*}
	\caption{CC-19 dataset collected from three different hospitals (A, B, and
		C).}
	\label{tab:dataset}
	
	\begin{tabular}{l>{\raggedright}p{0.1\textwidth}>{\raggedright}p{0.1\textwidth}>{\raggedright}p{0.1\textwidth}>{\raggedright}p{0.1\textwidth}>{\raggedright}p{0.1\textwidth}>{\raggedright}p{0.1\textwidth}}
		\toprule 
		Hospital ID & A & A & B & B & C & C\tabularnewline
		\midrule 
		CT scanner ID & 1 & 2 & 3 & 4 & 5 & 6\tabularnewline
		\midrule 
		Number of Patients & 30 & 10 & 13 & 7 & 20 & 9\tabularnewline
		\midrule 
		Infecation annotation & Voxel-level & Voxel-level & Voxel-level & Voxel-level & Voxel-level & Voxel-level\tabularnewline
		\midrule 
		CT scanner & SAMATOM scope & Samatom Definitation Edge & Brilliance 16P iCT & Brilliance iCT & Brilliance iCT & GE 16-slice CT scanner\tabularnewline
		\midrule 
		Lung Window level (LW) & -600 & -600 & -600 & -600 & -600 & -500\tabularnewline
		\midrule 
		Lung Window Witdh (WW) & 1200 & 1200 & 1600 & 1600 & 1600 & 1500\tabularnewline
		\midrule 
		Slice thickness (mm) & 5 & 5 & 5 & 5 & 5 & 5\tabularnewline
		\midrule 
		Slice increment (mm) & 5 & 5 & 5 & 5 & 5 & 5\tabularnewline
		\midrule 
		Collimation(mm) & 128{*}0.6 & 16{*}1.2 & 128{*}0.625 & 16{*}1.5 & 128{*}0.6 & 16{*}1.25\tabularnewline
		\midrule 
		Rotation time (second) & 1.2 & 1.0 & 0.938 & 1.5 & 1.0 & 1.75\tabularnewline
		\midrule 
		Pitch & 1.0 & 1.0 & 1.2 & 0.938 & 1.75 & 1.0\tabularnewline
		\midrule 
		Matrix & 512{*}512 & 512{*}512 & 512{*}512 & 512{*}512 & 512{*}512 & 512{*}512\tabularnewline
		\midrule 
		Tube Voltage (K vp) & 120 & 120 & 120 & 110 & 120 & 120\tabularnewline
		\bottomrule
	\end{tabular}
\end{table*}

\begin{table}[h]
	\begin{tabular}{|c|c|c|}
		\hline 
		S/No & Substance & Hounsfield Unit (HU)\tabularnewline
		\hline 
		1 & Air & -1000\tabularnewline
		\hline 
		2 & Bone & +700 to +3000\tabularnewline
		\hline 
		3 & Lungs & -500\tabularnewline
		\hline 
		4 & Water & 0\tabularnewline
		\hline 
		5 & Kidney & 30\tabularnewline
		\hline 
		6 & Blood & +30 to +45\tabularnewline
		\hline 
		7 & Grey matter & +37 to +45\tabularnewline
		\hline 
		8 & Liver & +40 to +60\tabularnewline
		\hline 
		9 & White matter & +20 to +30\tabularnewline
		\hline 
		10 & Muscle & +10 to +40\tabularnewline
		\hline 
		11 & Soft Tissue & +100 to +300\tabularnewline
		\hline 
		12 & Fat & -100 to -50\tabularnewline
		\hline 
		13 & Cerebrospinal fluid(CSF) & 15\tabularnewline
		\hline 
	\end{tabular}
	
	\caption{Various values of Hounsfield unit (HU) for different substances.\label{tab:hu}}
\end{table}

Collecting dataset is a challenging task as there are many ethical
and privacy concerns observed the hospitals and medical practitioners.
Keeping in view these norms, this dataset was collected in the earlier
days of the epidemic from various hospitals in Chengdu, the capital
city of Sichuan. Initially, the dataset was in an extremely raw form.
We preprocessed the data and found many discrepancies with most of
the collected CT scans. Finally, the CT scans, with discrepancies,
were discarded from the proposed dataset. All the CT scans are different
from each other i.e. CT scans have a different number of slices for
different patients. We believe that the possible reasons behind the
altering number of slices are the difference in height and body structure
of the patients. Moreover, upon inspecting various literature, we
found that the volume of the lungs of an adult female is, comparatively,
ten to twelve percent smaller than a male of the same height and age
\cite{DBLP:journals/AJRCCM/Bellemare}.

\subsection{Security Analysis's }

The use of permissioned blockchain distributed technology achieved
a secure mechanism for the various devices. We integrate the consensus
blockchain process with the federated learning to address the trust
of the security threats and privacy of the data.
\begin{enumerate}
	\item \textbf{To Achieve the Differential Privacy:} According to the privacy
	of users, our proposed protocol is used to indistinguishable for the
	random values. We select the random vector for generation of the ciphertext
	$\widetilde{c}_{i,}$ using the BGV scheme {[}31{]}. Where $K$ is
	indistinguishable security parameter for the random values. Then $v_{i}$
	is transfer from the polynomial $\widetilde{c}_{i,}$ for random values.
	\item \textbf{Data Access: }The proposed technique is used federated learning
	with blockchain technology, the core idea is to develop the privacy
	of the data. The proposed model achieves data privacy by aggregating
	the encrypted technique with blockchain, which grantee the privacy
	protection of the data
	\item \textbf{Aggregator the model trust security: }To aggregated sum of
	weights, the blockchain and local model client provide the security
	as fallows : 
	\begin{enumerate}
		\item Setup: First setup the security algorithm to generate the public parameters
		for the model 
		\item Encrypt: client specify the parameter $(i,m)$, Where $i$ is the
		index of the entity and $m$ is the plaintext. Finally, it returns
		the $Enc(i,m)$ value to the model.
		\item Compromise: The model comprises an $i$ entity, then aggregated model
		returns the secret keys $SK_{c}$ , this phase repeat many times.
		\item Challenge: It is allow only once throughout in the entire cycle. for
		every $i\in K$ generate and send two plain text $m_{1}$and $m_{2}$.
		If bit is equal to 0 then compute the $c_{i}={Enc}\left(m_{i}\right)$.
		Otherwise it will encrypted in the same way and send $c_{i}$
		\item Guess: The final output is 1 or 0 
	\end{enumerate}
	\item \textbf{Removing Centralized Trust: }The blockchain mechanism remove
	the third party trust and allows to users or hospitals commented with
	the decentralized network. 
	\item \textbf{Secure Data Management: }Only data trusted data provider upload
	the data to the network to ensure the reliability of the model. Moreover,
	the cryptography algorithm guarantee the security of the data.
	\item \textbf{Guarantee the Quality of Shared Model: }To prevent the quality
	of the model, consensus process guarantee the quality of learned data.
\end{enumerate}

\subsection{Performance Analysis }

\begin{figure*}
	\subfloat[]{\includegraphics[scale=0.34]{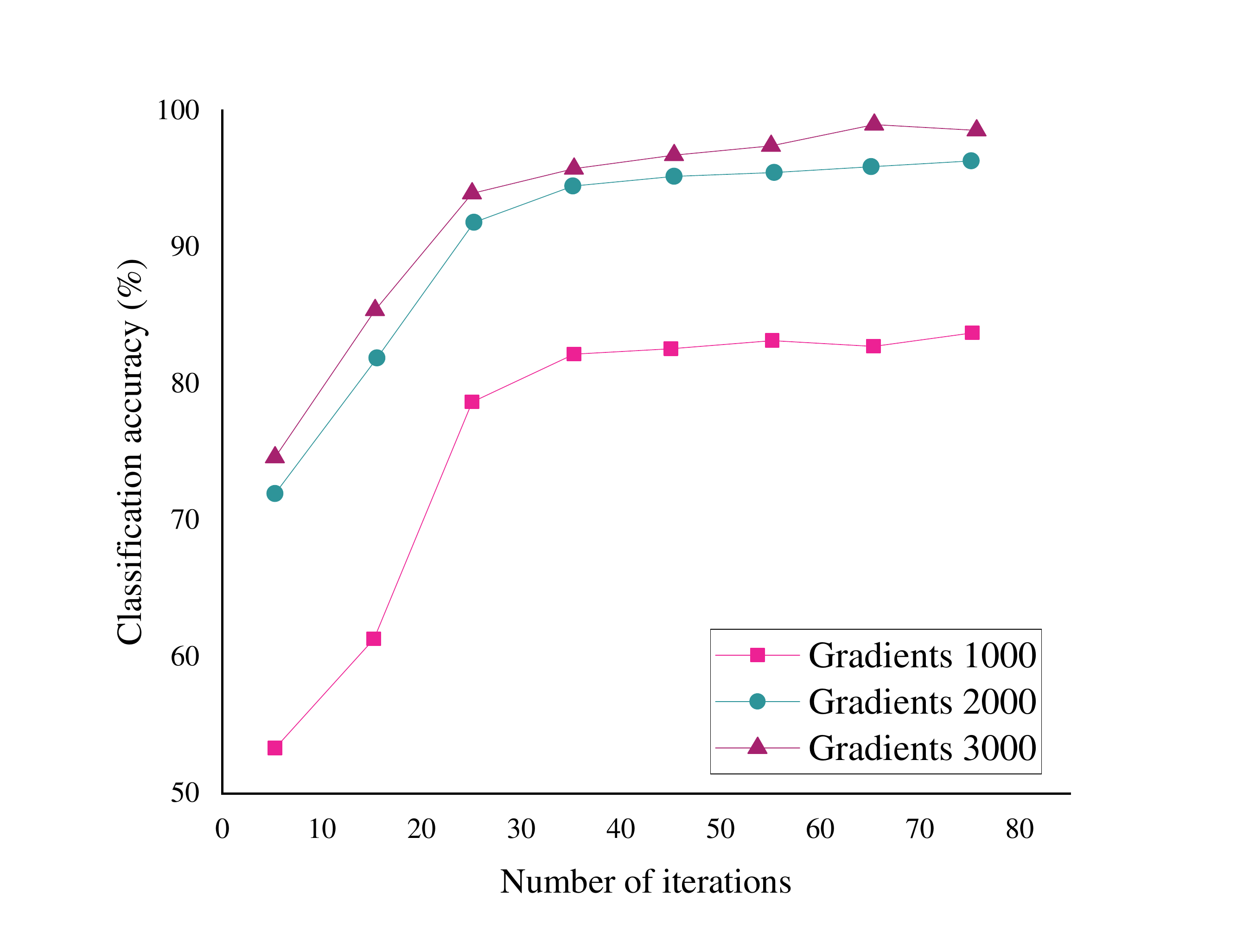}
		
		\label{Fig:gradient_with_accuracy}}\subfloat[]{\includegraphics[scale=0.34]{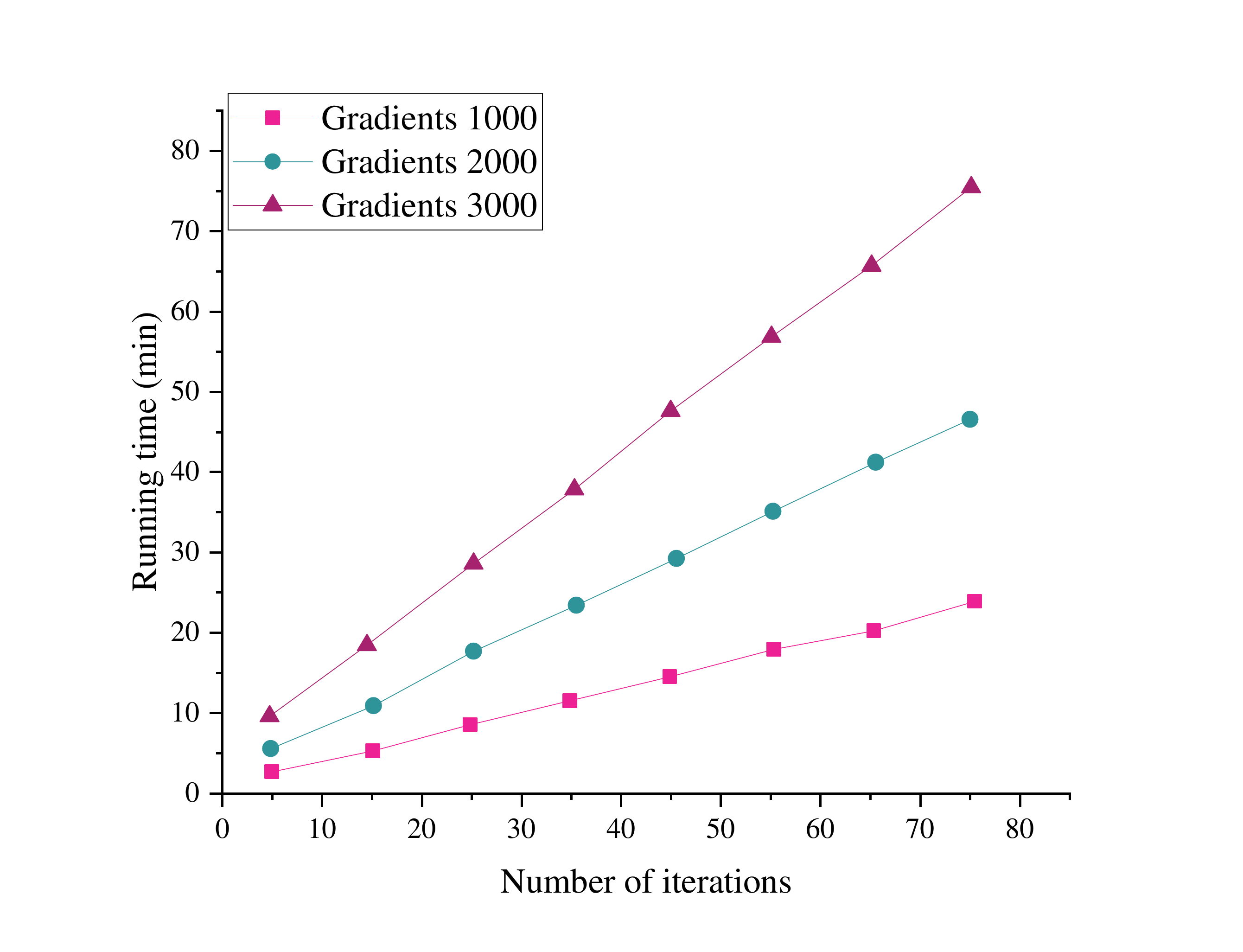}\label{Fig:gradient_with_running_time}}\caption{Hospitals=3 , no dropout, classification accuracy and running time
		for the various number of gradients per hospital.}
	\label{Fig:3}
\end{figure*}

\begin{figure*}
	\subfloat[]{\includegraphics[scale=0.34]{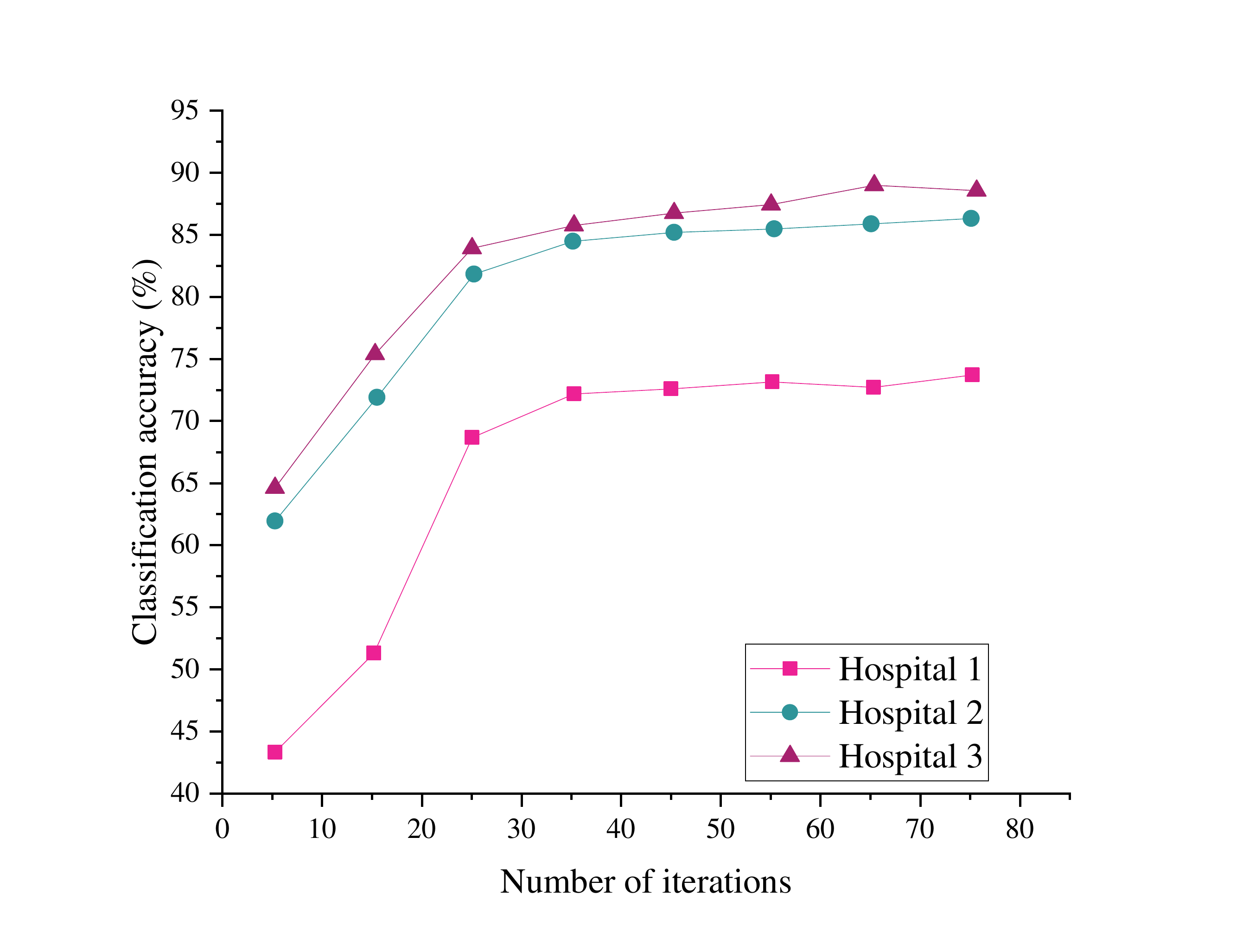}}\subfloat[]{\includegraphics[scale=0.34]{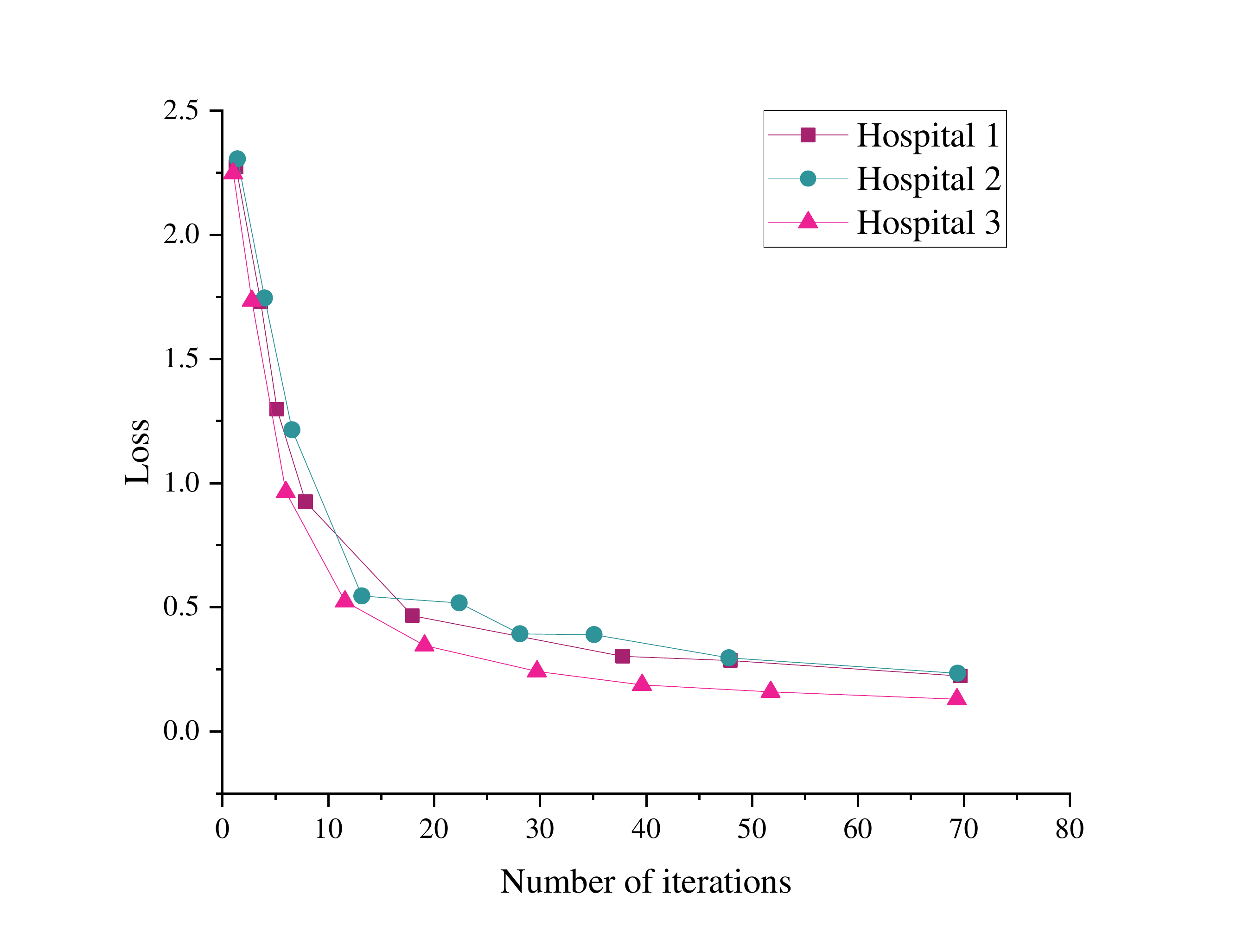}}\caption{Gradient=1000, no dropout, classification accuracy and loss for various
		numbers of hospitals.}
	\label{Fig:4}
\end{figure*}

To evaluate the proposed method's performance, we adopted the federated
learning model as a classifier to conduct the experimentation. We
analyze and evaluate our model in terms of accuracy. The deep learning
model; contains fully connected convolutional layers, where each of
the layers consists of 128 neurons. Two factors affect the accuracy
and running time of the federated learning model: the number of hospitals
and gradients per hospital. We analyzed both factors on different
ranges of values, as shown in Fig. \ref{Fig:3} and Fig. \ref{Fig:4},
respectively. \textcolor{black}{shows the execution time and accuracy
	with a different number of iteration. Here the number of iteration
	indicates the completed update of parameters. We compared the effect
	with the different number of gradients per hospital, and we distributed
	data to over six hospitals.  To conduct the experimentation on basic
	setting, we only assumed the condition where no user has dropped out.
	It can be clearly seen that increasing the number of gradients per
	hospital leads towards higher accuracy, whereas it causes the computation
	overhead as shown in Fig. \ref{Fig:gradient_with_running_time}.}
Therefore, to reduce the computation overhead in a practical environment,
an appropriate number of gradients can be empirically chosen. In terms
of model iterations, it can be observed that model accuracy converges
after a certain number of iterations. 

\textcolor{black}{The required time to train the local model (local
	gradients) also depends on the size of the data and the number of
	selected hospitals. We analyzed the accuracy over a different number
	of users to train the model. The classification accuracy and execution
	time can be seen in Fig. \ref{Fig:4}. Similar to the previous observation,
	naturally increasing numbers of iterations and hospitals consume high
	computation cost. However, due to independent gradient computation
	on each user, the number of hospitals leads to high accuracy. Basically,
	the data is split into many chunks as per hospital; therefore, the
	local gradient will be calculated and combined to produce high accuracy.}

\subsection{Local Model Capsule Network Performance Analysis }
In this section we analysis the local deep learning models which is divided into three parts (i) Segmentation (ii) Classification (iii)  Attention Map visualization

\subsubsection{Segmentation network results }

\begin{figure}[h]
	\centering \includegraphics[width=0.99\columnwidth]{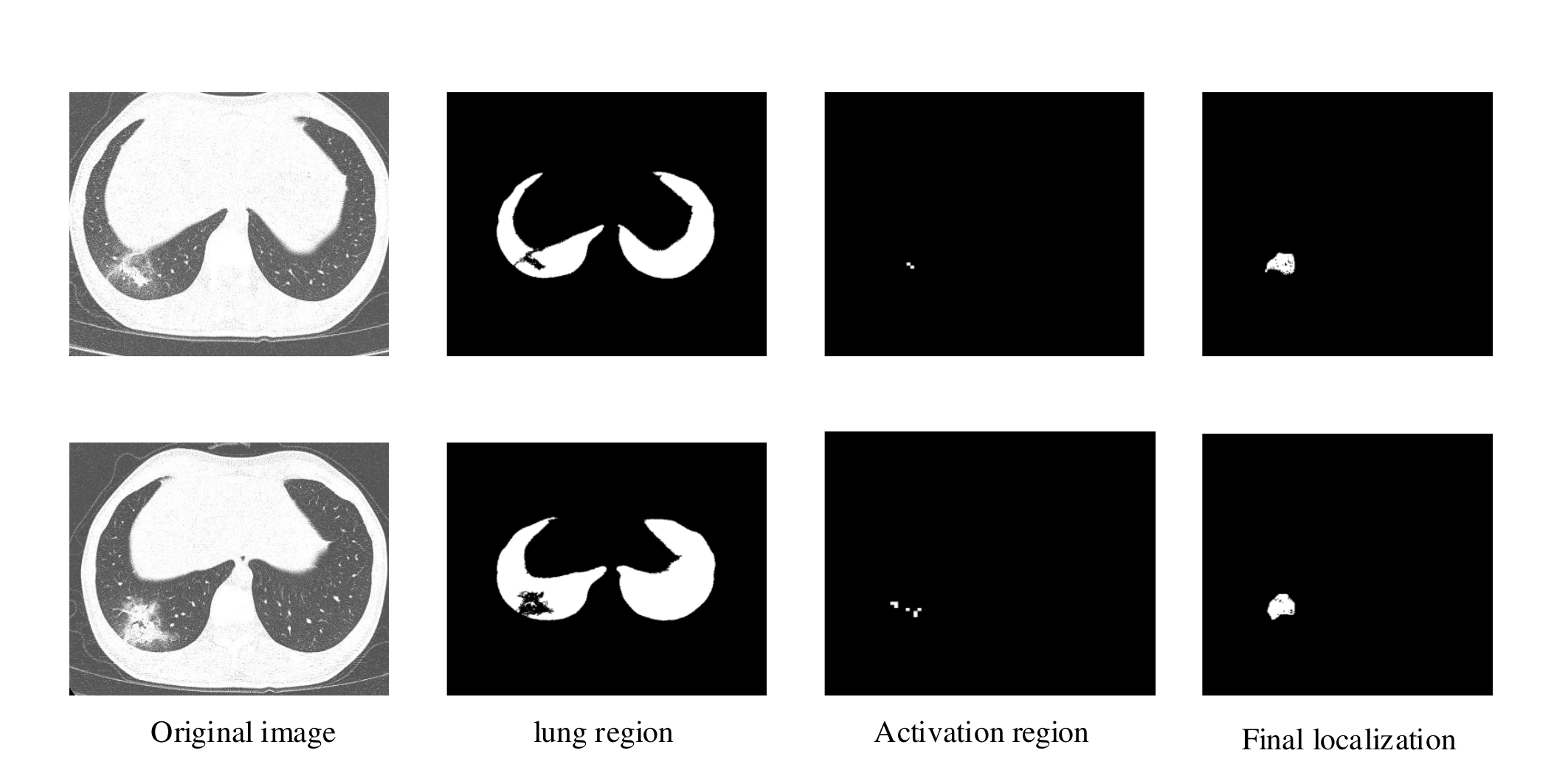}
	\caption{Activation mapping algorithm segmentation results}
	\label{fig:segmentation-results}
\end{figure}

Capsule network lesion localization of the lung's COVID-19 region
is shown in Figure \ref{fig:segmentation-results}. We extract the
region of the lung of COVID-19 patients. We fix the  parameters of the blockchain based federated learning, where total communication cost T to 300 and validate the each model in every round to select the best local model from the blockchain nodes. Moreover, we set the Adam optimizer learning rate of  0.0001. Each round contain 300 iteration with a batch size 4. Table \ref{tab:seg} shows the federated learning model for the three hospitals. First three rows  shows the (I/II/II) shows the hospitals. We compute the avrage of three hospital accuracy in  "global test avg". This measure shows the global model , and blockchain nodes as major metric for performance evaluation.       

\begin{table*}
	\centering
	\caption{COVID-19 lesion segmentation. Global test avg shows the Federated Learning global model.  $n$ spices the number of patients. }
	\label{tab:seg}
	
	\begin{tabular}{|c|c|c|c|c|c|c|}
		\hline 
		Parameters & Local - I & Local - II & Local III & FedAvg & FedAvg - Blockchain & FedProx\tabularnewline
		\hline 
		I ($n$=40) & 80.2 & 64.12 & 57.0 & 82.13 & 78.93 & 82.53\tabularnewline
		\hline 
		II ($n$=20)  & 84.02 & 82.15 & 74.74 & 85.99 & 86.51 & 87.18\tabularnewline
		\hline 
		III  ($n$=17) & 74.00 & 72.38 & 88.05 & 82.72 & 87.18 & 82.65\tabularnewline
		\hline 
		global test avg & 85.99 & 82.15 & 73.16 & 83.61 $\pm$0.18 & 84.21 $\pm$ 0.43 & 84.12 $\pm$ 0:58\tabularnewline
		\hline 
		local avg & \multicolumn{3}{c|}{84.07} & 84.67 & 84.44 & 61.99\tabularnewline
		\hline 
		local gen & \multicolumn{3}{c|}{70.99} & 81.0 & 81.48 & 80.53\tabularnewline
		\hline 
	\end{tabular}
\end{table*}

\subsubsection{Comparison the global and local  model}
\begin{figure}[h]
	\centering \includegraphics[width=0.99\columnwidth]{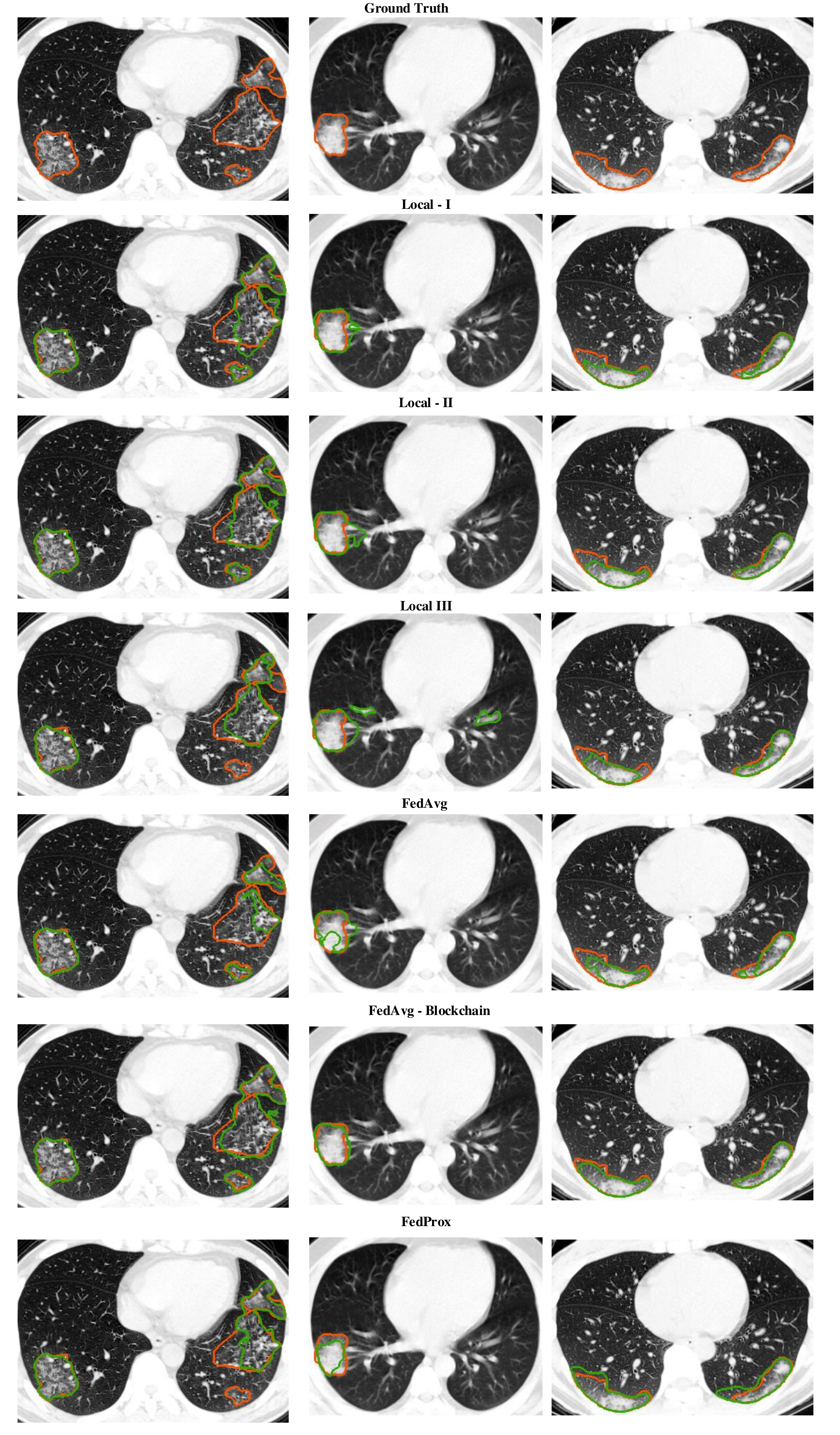}
	\caption{Activation mapping algorithm segmentation results}
	\label{fig:comapre-segmentation-results}
\end{figure}

This article conducts results from global and local  deep learning
models, i.e.,(Local I,  Local II, Local III, Fed AVG. Fed Global, FedProx). We used deep learning models and different layers
for comparing the performance models on the COVID-19 dataset, which
is shown in Figure \ref{fig:comapre-segmentation-results}. We evaluate
the performance of the capsule network for the detection of COVID-19
lung CT image accuracy. Figure \ref{fig:comapre-segmentation-results}
shows the local and global models; the global model achieves high high detection performance through the  network.  These
models were tested using three different test lists containing about
11,450 CT scan slices.

\subsubsection{Visualizations of the attention map regions}

\begin{figure}[h]
	\centering \includegraphics[scale=0.22]{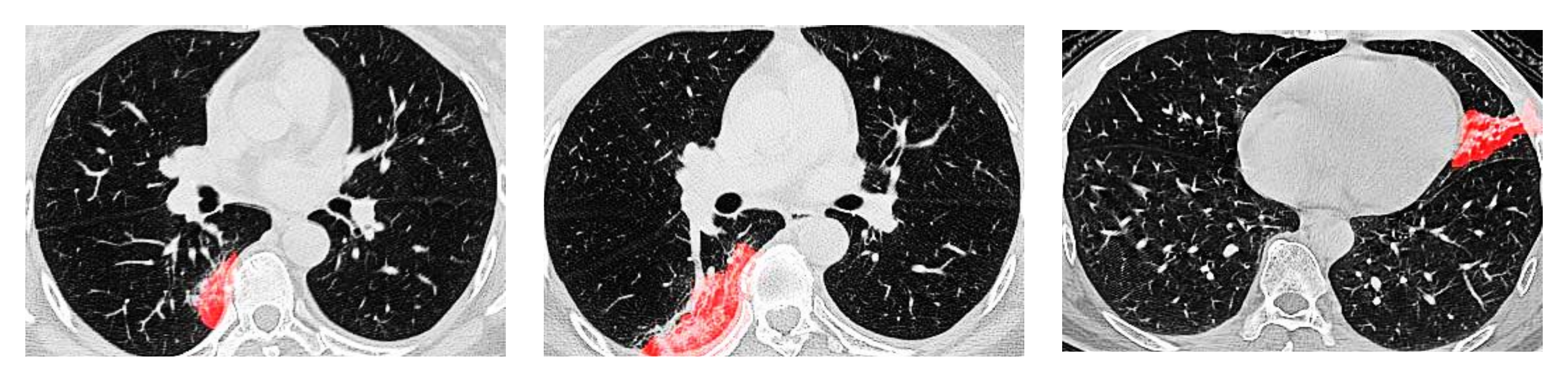}
	\caption{Visualizations of the attention map regions}
	\label{fig:attentation-map}
\end{figure}

To understand the deeper, we calculate the probabilistic CAM to each
CT image of COVID-19. The capsule network visualizes the patient CT
images from the normal and COVID-19 classes, and a noticeable activation
map is shown in Figure \ref{fig:attentation-map}. Moreover, the applied
CAM \cite{lalonde2018capsules,liao2019evaluate} function visualize
each image slice. These results strongly support our claim that the
probabilistic Grad-CAM saliency map. These results strongly support
our claim that the probabilistic Grad-CAM saliency map. 

\subsection{Compare with other methods }

To prove the local model accuracy and effectiveness of the proposed
model. As we can observe capsule network achieved 98\% accuracy in the detection
of the COVID-19 CT scans. Although Han el al. also achieve 98\% accuracy,
they do not consider the data sharing techniques. Furthermore, we
compare our scheme with the security analysis shown in Table \ref{Tab: Compare-security-analysis}.
However, Bonawitz et al. \cite{bonawitz2017practical}design a privacy-preserving
framework to secure the gradients' aggregation using the federated
learning global model. Zhang et al. \cite{zhang2017private} present
the homomorphic encryption (HE) scheme and threshold secret sharing
to secure the gradients. However, the shared model has no certainty
about authentic users. In other words, the trust issue between different
sources still exists; the proposed approach fill this gap and achieve
trust between parties.

\begin{table}[h]
	\begin{tabular}{|p{0.05\textwidth}|>{\centering}p{0.05\textwidth}|>{\centering}p{0.05\columnwidth}|>{\centering}p{0.05\textwidth}|>{\centering}p{0.05\textwidth}|>{\centering}p{0.05\textwidth}|>{\centering}p{0.05\textwidth}|}
		\hline 
		Study & Block- chain & Ser- ver & Data authentication & Privacy / Encryption Data & Data Access & Centra- lized Trust\tabularnewline
		\hline 
		\hline 
		OURs & Yes & No & Yes & Yes & Yes & Yes\tabularnewline
		\hline 
		\cite{kim2019blockchained} & Yes & No & Yes & No & Yes & Yes\tabularnewline
		\hline 
		\cite{Lu2020a} & Yes & No & Yes & No & Yes & Yes\tabularnewline
		\hline 
		\cite{Lu2020b} & Yes & No & Yes & No & Yes & Yes\tabularnewline
		\hline 
		\cite{xu2019verifynet} & No & Yes & No & Yes & Yes & No\tabularnewline
		\hline 
		\cite{yang2014secure} & No & Yes & No & Yes & Yes & No\tabularnewline
		\hline 
	\end{tabular}
	
	\caption{Compression with the security analysis }
	\label{Tab: Compare-security-analysis}
\end{table}

\section{Conclusion}

\label{sec:Conclusion}

This article proposed a secure data sharing scheme for the distributed
multiple hospitals for the internet of things applications, which
incorporate local model training and secure global training. We secure
the local model through the homomorphic encryption scheme, which helps
build an intelligent model without leakage the data provider's privacy
and create trust in the data training process. However, the blockchain-based
algorithm aggregates the local model updates and provides the authentication
of the data. The experiment results confirm the accuracy and effectiveness
of the model. In future work, to enhance the latency of the blockchain
and minimize the cost-effective solution.

\section*{Declaration of Competing Interest}
The authors declare that they have no known competing financial interests or personal relationships that could have appeared to influence the work reported in this paper.


\section*{Acknowledgement}
This work was sup­ported by Department of Science and Technology of Sichuan Province and National Science Foundation funding of china  , Project Number: H04W200533, U2033212.

\bibliographystyle{model2-names}\biboptions{authoryear}
\bibliography{ref}
\par\noindent 
\parbox[t]{\linewidth}{
	\noindent\parpic{\includegraphics[width=1in,height=1.25in,clip,keepaspectratio]{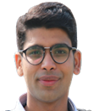}}
	\noindent {\bf Rajesh Kumar}\
	was born in Sindh Province of Pakistan in November 1991. He received his B.S. and M.S. degree in computer science from University of Sindh, Jamshoro, Pakistan. He received his Ph.D. in computer science and engineering from the University of Electronic Science and Technology of China (UESTC). He has currently Full time reseacher  in Yangtze Delta Region Institute (Huzhou), University of Electronic Science and Technology of China.  His research interests include machine learning, deep leaning, malware detection, Internet of Things (IoT) and blockchain technology. In addition, he has published more than 30 articles in various International journals and conference proceedings.}
\vspace{4\baselineskip}

\par\noindent 
\parbox[t]{\linewidth}{
	\noindent\parpic{\includegraphics[width=1in,height=1.25in,clip,keepaspectratio]{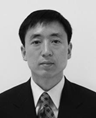}}
	\noindent {\bf Professor Wenyong Wang}\
	eceived the B.S. degree in computer science from Beihang University, Beijing, China, in 1988, and the M.S. and Ph.D. degrees from the University of Electronic Science and Technology (UESTC), Chengdu, China, in 1991 and 2011, respectively. He has been a Professor with the School of Computer Science and Engineering, UESTC, in 2009. He has served as the Director of the Information Center of UESTC and the Chairman of the UESTC-Dongguan Information Engineering Research Institute, from 2003 to 2009. He is currently a Visiting Professor with the Macau University of Technology. His main research interests include next-generation Internet, software-designed networks, software engineering, and artificial intelligence. He is a member of the expert board of CERNET and China Next-Generation Internet Committee and a Senior Member of the Chinese Computer Federation.}
\vspace{2\baselineskip}

\par\noindent 
\parbox[t]{\linewidth}{
	\noindent\parpic{\includegraphics[width=1in,height=1.25in,clip,keepaspectratio]{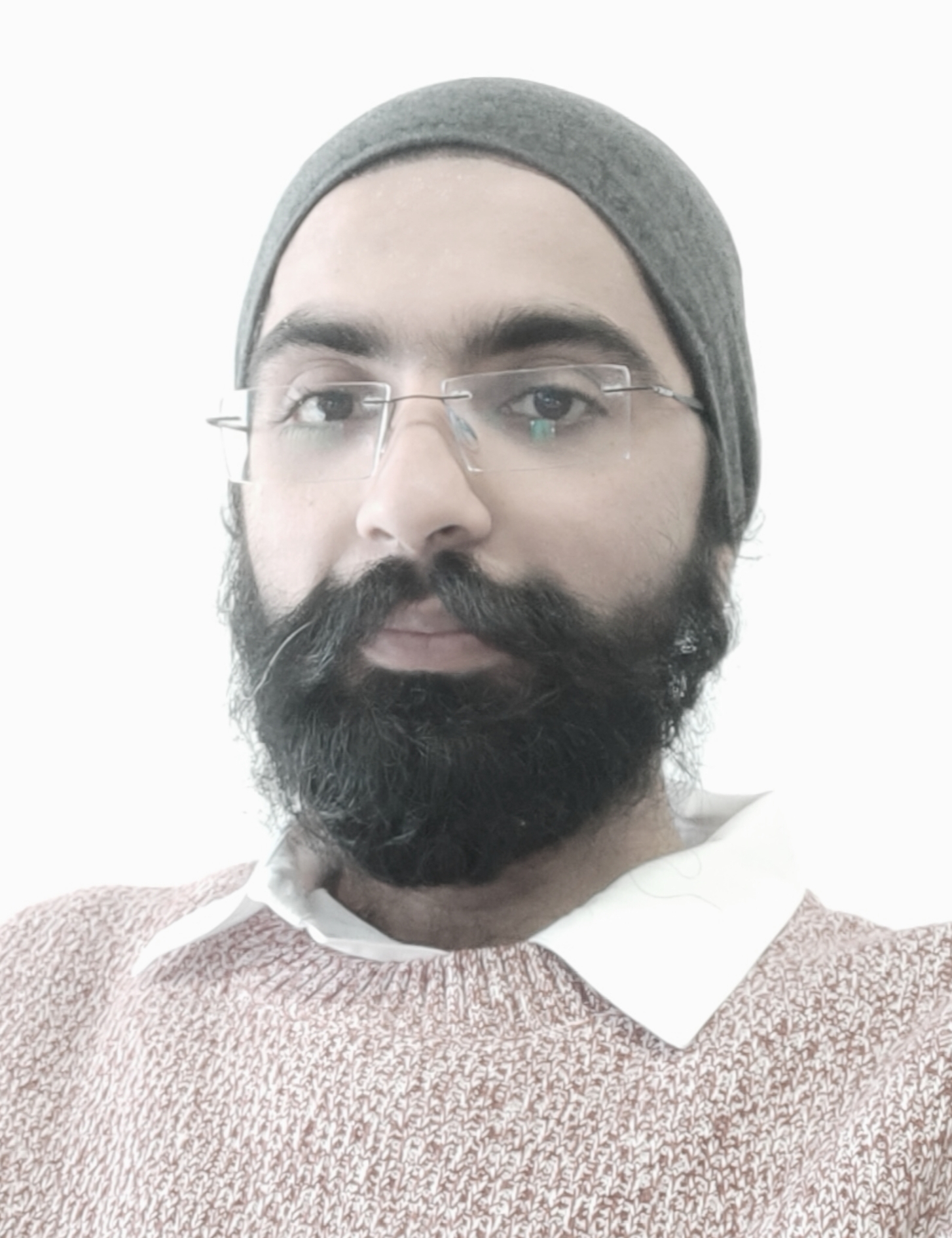}}
	\noindent {\bf Jay Kumar}\
	is currently a Ph.D student and working in Data Mining Lab, School of Computer Science and Engineering, University of Electronics Science and Technology of China. He received his Masters degree from Quaid-i-Azam University, Islamabad in 2018. His main interest of research include Text Mining, Data Stream Mining and Natural language processing. His current research work has been published in top conference of ACL, Journal of Information Sciences and IEEE transactions in Cybernetics.}
\vspace{2\baselineskip}

\par\noindent 
\parbox[t]{\linewidth}{
	\noindent\parpic{\includegraphics[width=1in,height=1.25in,clip,keepaspectratio]{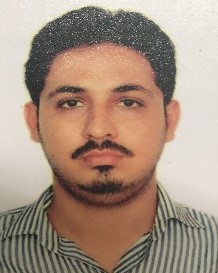}}
	\noindent {\bf Zakria}\
	received the M.S. degree in Computer Science and Information from N.E.D University in 2017. His Ph.D. degree with the School of Information and Software Engineering, University of Electronic Science and Technology of China. He has currently pursing Post Doctor in University of Electronic Science and Technology of China.  He has a vast academic, technical, and professional experience in Pakistan. His research interests include artificial intelligence, computer vision particularly vehicle re-identification.}
\vspace{2\baselineskip}

\par\noindent 
\parbox[t]{\linewidth}{
	\noindent\parpic{\includegraphics[width=1in,height=1.25in,clip,keepaspectratio]{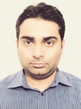}}
	\noindent {\bf Abdullah Aman Khan}\
	received his master’s degree in the field of Computer Engineering form National University of Science and Technology (NUST), Punjab, Pakistan in 2014. He is currently perusing PhD degree in the field of Computer Science and Technology form the school of computer science and engineering, University of Electronic Science and Technology, Sichuan, Chengdu, P.R. China. His main area of research includes electronics design, machine vision and intelligent systems.}
\vspace{2\baselineskip}

\par\noindent 
\parbox[t]{\linewidth}{
	\noindent\parpic{\includegraphics[width=1in,height=1.25in,clip,keepaspectratio]{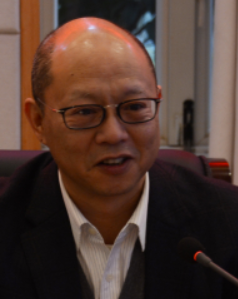}}
	\noindent {\bf Zheng Chengyu}\
	 is working as senior engineer and general manager of China Telecom Company Limited. He graduated from Huazhong University of Science and Technology, majoring in computer software design. He pursued executive master’s in business administration from Southwest University of Finance and Economics. He was selected by China Telecom Corporation to study at Stanford University in the United States. His main research directions include information technology system architecture design, mobile communication networks and buildin}
\vspace{2\baselineskip}

\end{document}